\documentclass[12pt,preprint]{aastex}

\accepted{}
\journalid{}{}
\articleid{}{}

\shortauthors{Mohanty et al..}
\shorttitle{Surface Gravities on Substellar Objects}

\received{2003 January}
\begin{document}

\def\tc{$T_{c}$~}
\def\tkh{$t_{KH}$~}
\def\hal{{H\alpha} }
\def\rc{R_C}
\def\ic{I_C}
\def\rcm{R_{Cm}}
\def\icm{I_{Cm}}
\def\rco{R_{Co}}
\def\ico{I_{Co}}
\def\av{A_V}
\def\ar{A_{R_C}}
\def\ai{A_{I_C}}
\def\kr{k_{R_C}}
\def\ki{k_{I_C}}
\def\rad{\mathcal{R}}
\def\dist{\mathcal{D}}
\def\mass{\mathcal{M}}
\def\lum{{\mathcal{L}}_{bol}}
\def\logten{log_{10}}
\def\na{\ion{Na}{1}\ }
\def\pot{\ion{K}{1}\ }
\def\hyd{\ion{H}{1}\ }
\def\hydmol{H$_2$\ }
\def\hel{\ion{He}{1}\ }
\def\kms{km s$^{-1}$\ }
\def\kmsp{km s$^{-1}$ pix$^{-1}$\ }
\def\cms{cm s$^{-1}$\ }
\def\cmc{cm$^{-3}$\ }
\def\cmss{cm$^{2}$ s$^{-1}$\ }
\def\cmcs{cm$^{3}$ s$^{-1}$\ }
\def\msun{M$_\odot$\ }
\def\rsun{R$_\odot$\ }
\def\lsun{L$_\odot$\ }
\def\mj{M$_J$\ }
\def\teff{T$_{e\! f\! f}$~}
\def\gv{{\it g}~}
\def\vsini{{\it v}~sin{\it i}~}
\def\vrad{v$_{\it rad}$~}
\def\lbol{L_{\it bol}}
\def\lhal{L_{H\alpha}}
\def\fhal{F_{H\alpha}}
\def\fbol{F_{\it bol}}
\def\lx{L_{X} }
\def\eqwhal{EW_{H\alpha}}
\def\alom{{\alpha}{\Omega} }
\def\ross{R_{0} }
\def\cots{{\tau}_{c} }
\def\fchal{{\mathcal{F}}_{c\hal} }
\def\h2o{H$_2$O}
\def\tross{${\tau}_{R}$ }

\title{Measuring Fundamental Parameters of Substellar Objects. I: Surface Gravities}

\author{Subhanjoy Mohanty\altaffilmark{1}, Gibor Basri\altaffilmark{1}, Ray Jayawardhana\altaffilmark{3}, \\France Allard\altaffilmark{4}, Peter Hauschildt\altaffilmark{5}, David Ardila\altaffilmark{6}}

\altaffiltext{1}{Harvard-Smithsonian Center for Astrophysics, Cambridge, MA 02138. smohanty@cfa.harvard.edu}
\altaffiltext{2}{Astronomy Department, University of California, Berkeley, CA 94720. basri@soleil.berkeley.edu}
\altaffiltext{3}{Astronomy Department, University of Michigan, Ann Arbor, MI 48109. rayjay@astro.lsa.umich.edu}
\altaffiltext{4}{CRAL, \'{E}cole Normale Sup\'{e}rieure, 46 Allee d'Italie, Lyon, 69364, France. fallard@ens-lyon.fr}
\altaffiltext{5}{Hamburger Sternwarte, Universitaet Hamburg, 21029 Hamburg, Germany. phauschildt@hs.uni-hamburg.de}
\altaffiltext{6}{Bloomberg Center for Physics \& Astronomy, Johns Hopkins University, Baltimore, MD 21218. ardila@adcam.pha.jhu.edu}

\begin{abstract} 
We present an analysis of high resolution optical spectra for a sample of very young, mid- to late M, low-mass stellar and substellar objects: 11 in the Upper Scorpius association, and 2 (GG Tau Ba and Bb) in the Taurus star-forming region.  Effective temperatures and surface gravities are derived from a multi-feature spectral analysis using TiO, \na and \pot, through comparison with the latest synthetic spectra. We show that these spectral diagnostics complement each other, removing degeneracies with temperature and gravity in the behavior of each.  In combination, they allow us to determine temperature to within 50K and gravity to within 0.25 dex, in very cool young objects.  Our high-resolution spectral analysis does not require extinction estimates.  Moreover, it yields temperatures and gravities {\it independent} of theoretical evolutionary models (though our estimates do depend on the synthetic spectral modeling).   We find that our gravities for most of the sample agree remarkably well with the isochrone predictions for the likely cluster ages.  However, discrepancies appear in our coolest targets: these appear to have significantly lower gravity (by upto 0.75 dex) than our hotter objects, even though our entire sample covers a relatively narrow range in effective temperature ($\sim$ 300K).  This drop in gravity is also implied by inter-comparisons of the data alone, without recourse to synthetic spectra.  We consider, and argue against, dust opacity, cool stellar spots or metallicity differences leading to the observed spectral effects; a real decline in gravity is strongly indicated.  Such gravity variations are contrary to the predictions of the evolutionary tracks, causing improbably low ages to be inferred from the tracks for our coolest targets.  Through a simple consideration of contraction timescales, we quantify the age errors introduced into the tracks through the particular choice of intial conditions, and demonstrate that they can be significant for low-mass objects that are only a few Myr old.  However, we also find that these errors appear insufficient to explain the magnitude of the age offsets in our lowest gravity targets.  We venture that our results may arise from evolutionary model uncertainties related to accretion, deuterium-burning and/or convection effects.  Finally, when combined with photometry and distance information, our technique for deriving surface gravities and effective temperatures provides a way of obtaining masses and radii for susbtellar objects independent of evolutionary models; radius and mass determinations are presented in Paper II.

\end{abstract}


\section{Introduction}
In the study of stars, there is perhaps no parameter more fundamental than stellar mass, which is pivotal in determining the entire evolutionary path traced by a star.  With the discovery of brown dwarfs, mass determination has become particularly important at the bottom of the Main Sequence.  After all, the very notion of brown dwarfs is predicated on mass:  these are substellar objects, which is to say they are less massive than the hydrogen-burning limit of $\sim$ 80 \mj. The derivation of ultra-low masses has become particularly crucial in the light of new claims of planetary mass objects (which we henceforth contract to ``planemos'') occuring in isolation in star-forming regions (e.g., \cite{Zapa00, Lucas00}).  While both planemos and brown dwarfs are substellar, the distinction between the two is drawn at the fusion boundary of $\sim$ 13 \mj: brown dwarfs undergo an initial phase of deuterium fusion, while planemos never harbor any fusion at all.  The existence of isolated planemos with a few Jupiter masses, if proven, has significant consequences for both star and planet formation (\cite{Padoan02, Bate02, Boss01}).  However, the key issue of whether the newly discovered free-floating objects really have planetary masses, or are simply misidentified brown dwarfs, remains unsettled for reasons we touch on below.  Only a precise mass derivation can unequivocally resolve this question.  

The empirical determination of substellar masses, though, has so far proven rather difficult.  The most direct approach is to apply Kepler's laws to binary (or higher-order) systems with known orbital parameters, in order to obtain a `dynamical mass'.  If the components can be directly observed, then the mass can be related to their other properties, such as temperature and luminosity, and the dynamical mass results extended to other directly detected bodies for which a dynamical mass cannot be acquired (e.g., free-floating objects).  However, suitable systems with brown dwarf components remain elusive (though a handful are now coming to light; e.g, \cite{Close02, Potter02, Lane01}).  Dynamical masses for planemos in circumstellar planetary systems (i.e., extrasolar planets) {\it have} been derived, but these objects have not been directly observed, so it is impossible to relate their masses to other physical properties\footnote{With the exception of the transiting planets HD 209458 and OGLE-TR056 (\cite{Charbonneau00} and \cite{Konacki03}, respectively), which have both mass and radius determinations.  Also, except in transiting cases, masses from RV surveys are only lower limits, due to the unknown inclination of the system.}.  To date, with the exception of HD 209458, no substellar object - whether brown dwarf or planemo - whose intrinsic spectrum has been observed has also been proven to be substellar by a direct, dynamical measurement of its mass.  Conversely, no object outside the Solar System with a dynamical substellar mass has actually been directly detected (again with the sole exception of HD 209458)\footnote{Some of the atmospheric features of the transiting planet HD 209458 have now been directly detected (\cite{Charbonneau02}; \cite{Vidal03}).  A comparison with the theoretical tracks reveals serious discrepancies between observed and predicted properties, probably due to stellar irradiation effects.  A total mass has now also been obtained for one binary system with a spectrum, GJ 569B, indicating at least one brown dwarf component; individual masses should be available in the near future; \cite{Lane01}.}.  Currently, the substellarity of all directly observed bodies is certified either through the lithium test, or by the fact that their effective temperatures are below the minimum Main Sequence temperature (\cite{Basri00}).  However, both tests can only indicate an upper limit to the mass, and not its precise value.  Moreover, they are not useful at very early ages, when lithium is undepleted even in low mass stars and substellar objects have not yet cooled sufficiently.

For these reasons, precise mass estimates for all directly observed substellar bodies have so far been obtained solely through comparison with theoretical evolutionary tracks.  However, the evolutionary models remain largely untested in an absolute sense, and different groups have generated somewhat different tracks (e.g., \cite{Baraffe98, Burrows97, Dantona94}).  Evolutionary modeling uncertainties arise from a variety of sources, such as the treatment of convection, choice of initial conditions and the modifying effects of any initial accretion phase.  Additionally, these uncertainties are exacerbated at the earliest ages and for very low masses (\cite{Baraffe02}).  It is precisely for these ages and masses, however, that the evolutionary models are currently most widely used: to distinguish between stellar and substellar objects in young clusters (since the lithium and temperature tests are not robust at these ages), as well as to infer the actual substellar masses.  For example, the planemo status of newly identified isolated bodies in young star-forming regions, mentioned earlier, depends entirely on the accuracy of the theoretical tracks.  Therefore, given the burgeoning uncertainty in the tracks with decreasing mass and age, it would be extremely useful to have an independent method for determining mass, in order to check the theoretical predictions.

We present such a method here, which relies on an accurate measurement of surface gravity from high-resolution spectra.  In a nutshell, we first derive surface gravities and effective temperatures (\teff), by comparing our high-resolution optical spectra to the latest synthetic spectra.  We then calculate radii and masses by combining our inferred gravity and \teff with photometry and distance measurements.  In this paper, we present the gravity and temperature spectral analysis.  The radius and mass calculations are presented in a companion paper (\citet{Mohanty03}; hereafter, Paper II). 

For cool, very low mass objects, the derivation of surface gravities from observed spectra is not trivial.  A vast multitude of molecular and atomic spectral lines arise in their low-temperature photospheres; an adequate modeling of these is a prerequite for inferring gravity (and \teff).  This has only become possible in the last several years, due to large advances in computing power and the compilation of increasingly comprehensive linelists.  The resulting synthetic spectra are now routinely used to infer the \teff and model the colors of low-mass stellar and substellar objects (e.g., \cite{Allard01, Schweitzer01, Leggett01, Leggett00, Burrows00, Basrimoh00, Kirk99}).  Here we take the next logical step, of using the synthetic spectra to derive surface gravities as well.  \teff and gravity, together with empirical distance and photometry information, then allow us to calculate masses and radii without resorting to evolutionary model predictions.

Clearly, the validity of our analysis depends on that of the synthetic spectra; we will assess the veracity of these spectra at appropriate junctures.  It must be pointed out, however, that synthetic spectra are essential even when masses are derived through comparisons with theoretical evolutionary models - either for translating observed spectral types to temperatures (when comparing the data to theoretical HR diagrams), or for converting theoretical temperatures and luminosities to expected photometry (when comparing the data to theoretical color-magnitude diagrams).  In other words, both our technique, as well as analyses using evolutionary models, are limited by any shortcomings in the synthetic spectra.  

In this paper we will derive, through spectral analysis, \teff and gravities for a sample of young, late-type (i.e., low mass) Pre-Main Sequence objects belonging to the Upper Scorpius and Taurus regions.  Since we observe them prior to their main contraction phase, we expect their surface gravities to be lower than in field objects of similar mass.  By comparing our results to the predictions of evolutionary models, we will also test these models in the low mass, young age regime where they are most uncertain.  

\section{Observational Sample and Properties}
Our sample consists of 11 mid- to late M objects in Upper Scorpius, and 2 in Taurus.  Our initial Upper Sco sample was culled from a study of this cluster by \cite{Ardila00} (hereafter, AMB00).  They chose objects that lie above the Leggett (1992) Main Sequence in both $I$ versus $R-I$ and $I$ versus $I-Z$ color-magnitude diagrams.  Their saturation limits also excluded all objects with $I \lesssim$ 13.5.  This procedure selects for pre-main sequence (PMS) objects roughly $\sim$ M2 and later, but is also subject to contamination by foreground low-mass stars, background giants and reddened background stars (AMB00).  Background galaxies are not expected to occur in the selected regions of the color-magnitude diagram (see AMB00 and references therein).  AMB00 then obtained low-resolution optical spectra for a subset of their sample.  In 20 of these, they detected $\hal$ in emission.  The strength of the emission was used as further (though not ironclad) evidence of cluster membership.

Here we explore the very low-mass regime in Upper Scorpius.  We originally included the 16 AMB00 objects with detected $\hal$ emission and spectral type M5 and later: these are the best candidates for lowest mass cluster members.  Of these, we were able to observe 13 on HIRES at Keck I.  We also observed an M7 Upper Scorpius candidate (USco 130) for which AMB00 had not obtained a low-resolution spectrum, bringing our total sample size to 14.  The high-resolution evidence for cluster membership and PMS status is discussed in \S 2.1.  We note in advance that, based on the latter evidence, 11 of the 14 objects observed appear to be bona-fide PMS members of Upper Scorpius.  These are listed in Table 1.  

Observations were made with the W. M. Keck I 10-m telescope on Mauna Kea using the HIRES echelle spectrometer (\cite{Vogt94}).  The observation particulars are listed in Table 1.  The settings used were similar to those described in \cite{Basmarcy95}.  The instrument yielded 15 spectral orders from 640nm to 860nm (with gaps between orders), detected with a Tektronix $2048^2$ CCD.  The CCD pixels are 24$\mu m$ in size and were binned 2 $\times$ 2; the bins are hereafter referred to as individual ``pixels''.  Each pixel covered 0.1A, and use of slit decker ``D1'' gave a slit width of 1.15 arcsec projected on the sky, corresponding to 2 pixels on the CCD  and a 2-pixel spectral resolution of R = 31000.  The slit length is 14 arcsec, allowing excellent sky subtraction.  The CCD exhibited a dark count of $\sim$2 e-/h and a readout noise of 5 e-/pixel. 

The data were reduced in a standard fashion for echelle spectra, using IDL procedures.  This includes flat-fielding with a quartz lamp, order definition and extraction, and sky subtraction. Dividing by the flat-field also corrects for the echelle blaze function\footnote{Although the slope of the quartz lamp is not exactly the same as the underlying blackbody part of the stellar slope, since the lamp and stellar temperatures are not identical, this difference is negligible over one echelle order for the very cool objects we are studying.  Dividing by the flat-field thus does not introduce any spurious slopes into our stellar spectra, while efficiently cancelling the contribution of the blaze function.}.  As we will show, the flat-fielded stellar spectra match the shape of model spectra very closely over a whole echelle order with no further corrections. The stellar slit function was found in the redmost order, and used to perform a weighted extraction in all orders.  The wavelength scale was established using a ThAr spectrum taken without moving the grating; the solution is a 2-D polynomial fit good to 0.3 pixels or better everywhere.  Cosmic-ray hits and other large noise-spikes were removed using an upward median correction routine, wherein points more than 7$\sigma$ above the median (calculated in 9-pixel bins) were discarded.

Barycentric and stellar radial velocity corrections were derived for each observed spectrum. We thank G. Marcy for the IDL routine to calculate barycentric corrections.  Radial velocities (\vrad) were derived by measuring the peak position for cross-correlation functions between each star and a \vrad standard. We chose Gl 406 (M6), with a known \vrad = 19$\pm$1 \kms (\cite{Martin97}), as our standard.  As a check on the accuracy of our measurements, we have calibrated our Gl 406 spectrum on an absolute wavelength scale.  We find its \vrad is indeed 19$\pm$1 \kms, in complete agreement with \cite{Martin97}.  

Our sample also includes GG Tau Ba and Bb, the two M-type components ($\sim$M6 and $\sim$M7.5 respectively) of the quadruple GG Tau system, located in the Taurus star-forming region.  Parameters for Ba and Bb have previously been derived through comparison with theoretical evolutionary tracks (White et al., 1999; hereafter WGRS99); it is thus instructive to reanalyse them using our technique, and compare our results to the previous ones.  High-resolution optical spectra of both objects were kindly supplied to us by R. White.  Like the Uper Sco spectra, these too were obtained with Keck HIRES, and reduced in a similar fashion by R. White.  

\subsection{PMS Status and Cluster Membership}
GG Tau Ba and Bb are already known to be bona-fide pre-Main Sequence (PMS) members of the Taurus complex (see WGRS99 and references therein).  Consequently, we focus here on the Upper Sco targets.  Optical high-resolution spectra yield four independent measures of PMS status and cluster membership: Lithium (LiI) detection, $\hal$ emission (already observed by AMB00 in our sample, but possibly overestimated due to blending effects at low-resolution), the width of gravity-sensitive lines (e.g., \na), and radial velocity measurements.  In this paper, we discuss only those objects that are PMS members of Upper Scorpius by {\it all four} criteria, i.e., those with detected LiI, clear $\hal$ emission at high-resolution, line widths intermediate between that of giants and field dwarfs, and \vrad consistent with cluster membership.  Below, we first discuss our observations of these four criteria, and then the reasons they robustly exclude all but PMS objects.

In 11 of our 14 objects, we detected LiI at 6708 \AA.  No LiI was seen in USco 85, 99 and 121; as a result, these are not likely to be PMS members of Upper Scorpius, and we do not discuss them further.  The LiI profiles are plotted in Fig. 1.  The corresponding equivalent widths have been quoted in \cite{jaya02} (hereafter, JMB02); they are mostly in the range $\sim$ 0.6 to 0.7 \AA.  In USco 128, we have a somewhat noisy LiI detection.  However, recent high-resolution spectra we obtained for this object, using the echelle spectrograph MIKE on the Magellan Baade telescope, clearly show LiI absorption.  The Magellan detection is plotted in Fig. 1; the corresponding equivalent width is 0.53\AA.

Our $\hal$ detections have been discussed, the observed equivalent widths quoted, and the $\hal$ line profiles shown in JMB02.  We only note here that, in the 11 objects with detected LiI, $\hal$ is strongly in emission, with equivalent widths ranging from $\sim$ 6 to 25 \AA.  The $\hal$ profiles and strengths are similar to those in dMe stars (see JMB02). 

The purpose of this paper is to derive stellar parameters such as gravity from an analysis of various atomic and molecular lines.  However, even a cursory examination can reveal if the observed spectra are commensurate with expectations for a low-mass, few Myr old PMS sample.  For such objects, theory predicts log \gv $\sim$ 3.0--4.0 (\cite{Chabrier00}).  M giants have much lower gravities (log \gv $\sim$ 0--2), while field M and L dwarfs exhibit substantially higher ones (log \gv $\sim$ 4.5--5.5).  In all our objects, the gravity sensitive \na doublet ($\sim$ 8200\AA) is narrower than in field M dwarfs and much broader than in M giants of similar spectral type, as indeed expected at an intermediate gravity.

Finally, while we do not have proper motions for our sample, we do have radial velocities.  The median \vrad for Upper Scorpius is -4.6 \kms (\cite{dezeeuw99} (hereafter ZHB99); the velocity dispersion is not cited).  We find -4.5$\pm$1 \kms for our sample, which is completely consistent with the ZHB99 value.  The Upper Scorpius OB association is a comoving group, so its members should have comparable velocities along all three spatial directions.  The excellent \vrad agreement thus makes it very likely that our objects are true members of this OB association.  We also point out that our \vrad rules out the possibility of interlopers from Upper Centaurus Lupus (UCL), another subgroup in Sco OB2 which somewhat overlaps the western edge of Upper Scorpius; the median \vrad in UCL is +4.9 \kms (ZHB99), well beyond our \vrad error bars.  We also note that USco 85, 99 and 121, which we do not consider to be Upper Scorpius members due to LiI non-detection, also differ in \vrad by $\gtrsim$ 15 \kms from our sample.  This further establishes them as non-members.

M dwarfs in the field do not exhibit Lithium in their spectra, having depleted it fairly early in their descent to the Main Sequence. The detection of LiI in our objects thus precludes their being field M dwarfs.  Field brown dwarfs (BDs) with mass $\lesssim$ 0.06 \msun do show Lithium, never having achieved core temperatures high enough to burn it.  However, such BDs have spectral types $\sim$ L2 and later, and log \gv $\approx$ 5-5.5 ($\gtrsim$ gravity in field M dwarfs), by young-disk ages ($\sim$ 1 Gyr).  Both the M-type classification of our objects, and their low gravity (as evidenced by their narrow \na profiles), argue against their being field brown dwarfs.

Lithium detection alone does not necessarily exclude Li-rich giants (though these are comparatively rare).  However, our \na line profiles clearly imply gravities higher than in giants.  Moreover, our detection of strong emission $\hal$ makes it very unlikely that our objects are isolated giants; $\hal$ in the latter is seen in {\it absorption} (at most, the line-core is somewhat filled-in by chromospheric emission, but the line as a whole is still in absorption).  Finally, while $\hal$ activity in giants can be somewhat enhanced when they are a part of Algol-type or RS CVn binaries (\cite{richards99}; \cite{fox94}), the giants in these cases are always in the F - K spectral type range; our objects are mid- to late M.  Reddening cannot lead us to infer spuriously late spectral types either: spectral types for our objects have been determined by AMB00 from low-resolution spectra, not colors, and our high-resolution observations imply M-types as well (e.g., through the presence of strong TiO bands)\footnote{Usco 130 is the only exception, in that AMB00 do not have a spectrum for it, and derive a spectral type from colors alone.  However, our high-resolution spectrum confirms AMB00's late-M classification of this object, and precludes any misidentification due to reddening effects.}.  In summary, the spectral type, LiI detection, dMe-like $\hal$ emission and (relatively) higher gravity of our objects all argue strongly against any of them being giants.  

Finally, our \na profiles are commensurate with sub-giant gravities, as indeed expected for PMS objects.  However, it is extremely unlikely that any of our objects are sub-giant interlopers.  The most convincing argument for this is simply that M type sub-giants are not expected to exist; the MS lifetime of M dwarfs is too long for any to have reached the sub-giant stage yet.  Additionally, the arguments made above, concerning the unlikelihood of LiI absorption and dMe-like $\hal$ emission in giants, hold for sub-giants as well.  The PMS status of our final Upper Sco targets is thus very secure; moreover, their radial velocities suggest that they are very likely to be bona-fide members of the Upper Scorpius association.  

\subsection{Rotation Velocity}
Our method for deriving projected rotation velocities (\vsini) is discussed in detail in \cite{Mohanty02}.  Briefly, we find \vsini through cross-correlating our target spectra with a template spectrum.  We use only those echelle orders without emission features, broad stellar absorption lines and telluric features.  Cross-correlation functions are found in multiple orders ($\sim$ 6), and the average is used to determine \vsini.  \cite{White02} find that for very young objects, a template constructed by averaging a dwarf and giant spectrum is appropriate for spectral type $\sim$ M6 and later, while a dwarf spectrum alone is more suitable for earlier types.  Moreover, they find that for their sample of M type Taurus objects, cross-correlating with an M giant spectrum yields \vsini lower by $\sim$ 8-9 \kms than cross-correlating with an M dwarf spectrum.  This is because the molecular lines in the giant appear somewhat broader than in the dwarf, even though the giant is at lower gravity, presumably because of turbulence and/or NLTE effects\footnote{The alternative explanation, that the giants are actually rotating faster than the dwarf, is very unlikely, as \cite{White02} point out; if the broadening were due to rotation, the giants observed by \cite{White02} would all have \vsini $\sim$ 10 \kms, which is highly uncommon for such evolved stars.}.  Consequently, cross-correlating with a giant-dwarf averaged template yields intermediate \vsini values: $\sim$ 4-5 \kms higher than with a giant alone, and $\sim$ 3 \kms lower than with a dwarf (\cite{White02}).

Our sample consists of M5-M8 objects, with 4 in the M5-5.5 range.  Gl 406 (M6, \vsini $<$ 3 \kms (\cite{Mohanty02})) has been used as our dwarf template, and HD 78712 (M6IIIe, presumed non-rotating (\cite{White02})) as our giant template.  For our sample, our analysis agrees with that of \cite{White02}, with one additional result.  For objects with \vsini $\gtrsim$ 10 \kms, the \vsini obtained through cross-correlation with the giant-dwarf averaged template differs by $\lesssim$ 2 \kms from the \vsini obtained using either the dwarf or giant template alone.  That is, at moderate to high rotation rates, the difference in \vsini implied by the three templates is small; this is to be expected, as rotational broadening overwhelms other broadening mechanisms with increasing \vsini (this result is moot to the \cite{White02} work, since all but 2 of their Taurus objects have \vsini $\lesssim$ 10 \kms).  At the same time, our precision in determining \vsini, for any given template, is of order $\pm$ 2 \kms (see \cite{Mohanty02} for detailed discussion).  Now, we find \vsini $>$ 10 \kms in all but 5 of our objects (regardless of whether the giant, dwarf or averaged template is used), and $>$ 15 \kms in most.  Therefore systematic errors in our \vsini due to any template mismatch are small, and in most cases less than our random errors.  Our final \vsini values are given in Table 1; the Upper Sco ones have also been quoted earlier in JMB02.  We note that the latter paper gives 5.5$\pm$2 \kms for USco 112, while here we find 8$\pm$2 \kms.  This is because all values in the former paper were derived using only giant-dwarf averaged templates, including in the M5-5.5 objects.  Here, we use dwarf templates for these spectral types; since \vsini for this object is $<$ 10 \kms, the template change makes a (small) difference.  All our other Upper Sco targets remain unaffected, for the reasons cited above.  We also note that, due to instrumental broadening effects (see \cite{Mohanty02} for details), our lower limit for \vsini detection is $\sim$ 5 \kms.  

\section{Surface Gravity \& Effective Temperature Analysis}
Having established the PMS nature of our sample, and derived the projected rotational velocities, we are now ready to determine \teff and gravity.  The synthetic spectra to which our data are compared are described in \S 3.1, followed by a discussion of the modifications made to the models and data, and of our fitting procedure, in \S 3.2.  	Our method of analysis is detailed in \S 3.3.  We present the results in \S 4, and the attendant implications in \S 5. 

\subsection{Model Spectra}
We use the latest version of synthetic spectra generated with the PHOENIX code (\cite{Allard01}; Allard et al., in prep.).  Specifically, we use the models designated AMES-Cond-2002-version2.4.  These incorporate the most recent AMES line-lists for both TiO (\cite{Langhoff97, Schwenke98}) and \h2o (\cite{Partridge97}).  A good treatment of \h2o is important for analysing optical spectra, even though \h2o opacity only dominates in the infrared (TiO opacity is more important in the optical).  This is because the overall \h2o opacity is larger, and its lines occur closer to the peak of the SED than those of TiO, at the low \teff in M types.  Consequently, changes in the \h2o opacity have a substantial effect on the atmospheric temperature structure, and thus on the emergent spectrum even in the optical.  A total of about 500 million molecular lines are currently included in the models; of these, $\sim$ 307 million are lines of \h2o, and $\sim$ 172 million are of TiO (\cite{Allard00}; \cite{Allard01}).  In this paper, we only use solar-metallicity models ([M/H] = 0.0).  While metallicity in the Upper Sco cluster has not yet been rigorously investigated, large deviations from solar are unlikely for nearby, young PMS objects (see also \S 4.3.3).

Dust formation is another potentially important effect in the low temperature atmospheres of M type objects; grains affect both the atmospheric structure and the emergent flux.  Current models include these effects.  The `Dusty' models treat the case where grains form and remain suspended in the photosphere, while the `Cond' models are applicable when dust has settled (or `condensed') out of the atmosphere.  Both sets of models treat grain formation self-consistently, through chemical equilibrium calculations (see \cite{Allard01}).  It is important to note, however, that under physical conditions where the chemical equations imply {\it no} grain formation, the `Cond' and `Dusty' models produce identical spectra: when no grains form, the treatment of grain settling or suspension is immaterial.  

In field dwarfs, spectral analyses indicate that dust formation becomes important around M7 (\cite{Jones97}), and dust settling at still later spectral types, $\sim$ L2 (e.g., \cite{Schweitzer01}).  In lower gravity PMS objects like ours, no detailed empirical spectral analysis of photospheric dust has yet been carried out.  Nevertheless, since the latest spectral types in our sample are $\sim$ M7-7.5, it is possible, by analogy with field dwarfs, that dust begins to appear in their photospheres.  To ascertain whether significant grain formation occurs in our objects, we have performed two tests: the first checks if the synthetic spectra imply any photospheric dust in our sample, and the second examines whether our observed spectra show any empirical signatures of dust (independent of the synthetic spectra).

In the first test, we compare the spectra of our latest type objects to both the `Dusty' and `Cond' models.  We find that the two yield identical \teff and gravities; i.e., for the best fit \teff and log \gv, the models indicate no grain formation.  In the spectral regions used in our \teff and gravity analysis (see \S 3.3), grain effects first appear weakly in the models at $\sim$ 2500K, and strengthen with decreasing \teff.  The lowest \teff we derive, on the other hand, with either `Cond' or `Dusty', are $\sim$ 100K higher. We have also examined whether the combined effects of \teff and dust can create misleading degeneracies; i.e., whether higher \teff models without grain formation can mimic lower \teff models with grains, yielding erroneous temperatures (and gravities).  No degeneracies pertinent to our analysis are found.  In the second test, we compare our observed PMS spectra to those of field M dwarfs of similar spectral type.  This allows us to empirically gauge if spectral signatures of photospheric grains, which are evident in the field dwarfs, arise in our sample as well. The detailed analysis is presented in Appendix A and summarized in \S 4.3.1; here, we simply state our conclusion that grain formation is negligible in our PMS sample.

In summary, our tests show that neither the synthetic spectra, nor the observations, imply any significant grain effects in our targets.  Under the circumstances, we will show only `Cond' models in this paper; this is simply because, at the time of writing, these models were available to us over a larger range of \teff than the `Dusty' ones\footnote{This does not affect our aforementioned comparison between the `Cond' and `Dusty' models to check for dust formation: the comparisons have been carried out at the low-\teff (later spectral type) range of our sample, where we have both sets of models; since the models imply no dust formation (ie, `Cond' and `Dusty' are indentical) at these \teff, we can be confident that no dust appears in the models at higher \teff either, where we have only `Cond' models, and not `Dusty' ones.}.  Once again, we emphasize that the `Cond' appellation does {\it not} imply any grain formation or settling in our objects: fits to either `Cond' or `Dusty' models yield the same \teff and gravities.  

The treatment of convection is also important for our sample.  Currently available synthetic spectra and evolutionary models use a mixing length formulation, which seems to reasonably approximate the true convective transport mechanism (see \cite{Chabrier00} for a full discussion).  In certain regimes, the actual value of the mixing length parameter ($\alpha$) that one adopts turns out to be unimportant, for the spectral and evolutionary modeling.  For instance, for more evolved (i.e., higher gravity) low-mass ($\lesssim$ 0.6 \msun) objects, varying $\alpha$ from 1 to 2 hardly affects the models.  However, as \cite{Baraffe02} point out, this is {\it not} true for the same low-mass objects during their PMS phase.  At the low gravities associated with the initial contraction phase, changing $\alpha$ can affect the dissociation of \hydmol in the deeper atmospheric layers, and hence the \hydmol opacity, atmospheric structure and emergent spectrum.  Earlier generations of synthetic spectra, as well as some of the most widely used current evolutionary models, use $\alpha$=1 in the PMS, low-mass stellar and substellar regime.  However, full 3-D hydrodynamical simulations carried out very recently indicate that $\alpha$ $\approx$ 2 is a better approximation for mid-M types, both in the field as well as in the PMS phase (\cite{Ludwig03}).  Such simulations offer the best insight so far into the actual convection process (see \cite{Chabrier00}); consequently, the latest synthetic spectra we use here incorporate $\alpha$=2.  We will address the mixing length issue further, when we contrast our results with the evolutionary track predictions.

Finally, the treatment of collisional pressure broadening of the alakali lines is important for our analysis, since we shall use these line profiles to derive gravities.  A full exposition of the line-broadening treatment in the synthetic spectra is given by Schweitzer (1995); a summary can be found in Schweitzer et al. (1996).  We briefly discuss broadening in \S 4.3.4.
  
\subsection{Modifications to Models and Data, and Fitting Procedure}
In order to compare the synthetic spectra to the data, some modifications are necessary to both.  The observed spectra are rotationally broadened; to simulate this, the models have been convolved with the appropriate velocity profiles (depending on the \vsini we derive independently of the models; see \S 2.2).  Moreover, the data have a finite instrumental resolution.  The models have therefore been further broadened, by convolution with a Gaussian profile, to match our HIRES resolution of 31,000.  

Our high-resolution data are not flux-calibrated.  Comparison to the models thus requires some form of normalization.  As described in the next section, we will analyse various spectral intervals, containing TiO bandheads and alkali doublets.  We have normalized the data in each of these intervals by a constant factor.  The latter is the mean of the continuum over a small range of wavelengths; the range chosen depends on the echelle order.  Note that in these cool objects, there is no `true' continuum, only a {\it pseudo}-continuum made up of millions of overlapping molecular lines; however, for brevity, we drop the `pseudo-' appelation from here on.  The models are normalized by the model-continuum mean over the same wavelength ranges, after correcting for radial velocity shifts relative to the data.  Telluric lines, strong stellar lines that appear or disappear with changing \teff or gravity, or strong stellar lines that are not well-reproduced in the synthetic spectra, can all potentially skew the mean.  The normalization wavelength ranges have been chosen, through intercomparison of all our data and models at different \teff and gravities, to specifically exclude such features.  The spectra are checked during fitting to ensure that the normalizations are indeed correct, and not thrown off by, say, the appearance of random noise spikes in the data within the normalization wavelength range.

Recall that the data have also been flat-fielded, which removes the blaze-function and reveals the intrinsic shape and slope of the continuum.  Our normalization method then ensures that this innate stellar continuum, shaped by various opacity effects, remains undisturbed.  That is, we have not attempted to normalize out the intrinsic appearance of the stellar continuum, in either the data or the models, through any further division by an overall fit to the continuum.  In any given spectral region of interest, the data and models are only `anchored' over a small wavelength region - i.e., are scaled to ensure only that their continuua overlap over this section - and are otherwise unconstrained.  This allows us to compare the observed and predicted continuum behavior, and provides an additional check on the veracity of the synthetic spectra, as described shortly (\S 3.3).
  
Finally, the models were intially constructed in steps of 100K in \teff, and 0.5 dex in gravity.  We will demonstrate in \S 4.1 that the TiO bandheads and alkali doublets used in our analysis actually appear sensitive to even smaller changes in temperature and gravity.  We have thus averaged models separated by 100K to construct a model grid with 50K resolution in \teff; analogous averaging of models separated by 0.5 dex yields a grid with 0.25 dex resolution in log \gv (all averaging is done after normalizing the models in the manner described above).  Such a procedure is justified: a perusal of the normalized synthetic spectra reveals that our spectral diagnostics - TiO bandhead strengths, alkali doublet line profiles, and shape and slope of the continuua - all vary fairly linearly over $\sim$ 200K changes in \teff and$\sim$ 1 dex changes in gravity, over the entire \teff and gravity range we will consider (\teff = 2500-3000K, log \gv = 3.0-4.5).  

Best fits to the synthetic spectra were intially found through visual inspection.  In \S 4.1 we show that \teff and log \gv mismatches of 50K and 0.25 dex respectively, between the models and data, are distinguishable by eye; mismatches at the 100K and 0.50 dex level are obvious.  We conservatively estimate errors of $\pm$ 50K and $\pm$ 0.25 dex (see \S 3.3 and 4.1).  To check our results, we have also derived chi-squared differences between the models and data (after excluding regions with telluric contamination), over a $\pm$100K (in steps of 50K) and $\pm$0.50 dex (in steps of 0.25 dex) range centered on the best-fit \teff and log \gv values derived by eye.  In all cases our initial results, and our estimated errors, are validated:  the chi-square values are always minimum for the visually identified best-fit model, models within 50K and 0.25 dex of the best-fit one yield somewhat higher chi-squares, while models differing by 100K or 0.50 dex from the best-fit one produce much larger chi-square values.

\subsection{Method of Analysis}
In order to determine \teff and gravity, we compare the observed high-resolution line profiles of various temperature- and gravity-sensitive species to synthetic spectra.  Specifically, we examine:
\begin{enumerate}
\item The highly temperature-sensitive triple-headed band of Titanium oxide (TiO) at $\lambda\lambda$8432, 8442, 8452 \AA;
\item The gravity- and temperature-sensitive doublets of the neutral atomic alkalis Sodium (\na; subordinate doublet at $\lambda\lambda$8183, 8195 \AA) and Potassium (\pot; resonance doublet at $\lambda\lambda$7665, 7699 \AA);
\item The overall shape and slope of the continuum over the entire HIRES orders in which the \na and \pot doublets are located; each order spans $\sim$ 100 \AA.
\end{enumerate}

In M types, TiO is the main source of optical continuum opacity\footnote{Vanadium oxide (VO) also contributes significantly to the continuum opacity, though less than TiO.  As noted previously, there is only a pseudo-continuum in these objects, comprising millions of weak, overlapping molecular lines.  In the optical, most of these are lines of TiO, and to a lesser degree, VO.}.  The strengths of many of the bandheads of this molecule are very dependent on temperature (see \cite{Solf78} for detailed discussion).  We make use of the triple-headed band at $\lambda\lambda$8432, 8442, 8452, identified as the electronic E$^3$$\Pi$ - X$^3$$\Delta$ system (\cite{Solf78}).  Its response to changes in \teff and gravity is depicted in Fig. 2, using the synthetic spectra.

It is clear that this TiO system is highly sensitive to \teff (increasing in strength with decreasing \teff), and rather {\it in}sensitive to variations in gravity.  At a given \teff, the band-strengths change only slightly even for a large, 1 dex change in gravity (in the log \gv $\sim$ 3.0-4.0 range expected for low-mass PMS objects).  At a given gravity, however, they vary significantly over only 100K changes in \teff (specifically, the bandheads at 8442 and 8452 \AA; larger changes in \teff similarly affect the bandhead at 8432 \AA~ as well).  Indeed, as we will show later in comparisons to actual data, these bands enable us to distingish \teff variations down to $\pm$ 50K.  This TiO system is thus an excellent temperature indicator.  
  
However, since our goal is to fix gravity as accurately as possible, we cannot altogether ignore the small gravity-dependence of these TiO bands.  Fig. 2 shows\footnote{In this and following figures, the cited spectral differences may not always be completely obvious, given the reduced plot-scale we have had to adopt in the interests of space.  However, plots in the electronic copy of the journal can be enlarged; the interested reader is requested to consult these.} that at such temperatures, increasing gravity (at a fixed \teff) makes the 8442 and 8452 \AA~ bandheads slightly weaker, while decreasing gravity makes them a little stronger\footnote{The 8432 \AA~ bandstrength is clearly even less affected by gravity variations; we {\it will} ignore its \gv-dependence henceforth, since the dependence is much too small to affect, or be useful for, our analysis.}.  As a result a (large) change in gravity can compensate for a (small) change in \teff, and neither quantity can be accurately fixed.  Of course, a believable range in gravities and \teff can still be assigned by appealing to spectral type and/or theoretical evolutionary track considerations.  However, given our stated purpose of {\it testing} the latter, this is unsatisfactory.  What is needed is an additional spectral constraint on the TiO results.

This is provided by the \na and \pot alkali doublets, whose profiles also depend on both \teff and gravity.  Let us consider \na first.  It's variation with \teff and gravity is depicted in Fig. 2.  We see that the doublet strength increases with both decreasing \teff and increasing gravity.  For example, either a 200K decrease in \teff {\it or} a 0.5 dex increase in log \gv changes the line-profile similarly, making the line core slightly deeper and the line wings substantially broader\footnote{We note that at the \teff and gravities under consideration - 3000-2500K and 3.0-4.0 dex - the \na and \pot doublets are unsaturated.  In Fig. 2, this is evinced for \na by the slightly unequal core-depths in the two components of the doublet.}(Fig. 2).  Indeed, the shape of the \na doublet changes appreciably even for a 0.25 dex change in gravity (or equivalently, a 100K change in \teff).  Like for TiO, this degeneracy between \teff and gravity precludes our using the \na doublet, in isolation, to uniquely determine either quantity.

When considered together, however, TiO and \na can fix both \teff and gravity very precisely.  To see this, notice that the \teff-gravity degeneracies act in opposite senses in TiO and \na.  That is, while decreasing \teff compensates for increasing gravity in the TiO bands considered here, an {\it increase} in \teff offsets a gravity increase in \na.  This behaviour allows us to quickly converge on the correct \teff and gravity, if we demand that TiO and \na fits be obtained at the same temperature and gravity.  Deviations from the true gravity (or \teff) leads to large discrepancies in the \teff (or gravity) implied by TiO and \na.  

An example of this fitting procedure is illustrated in Fig. 2.  Suppose an object has \teff=2800K and log \gv=3.5 (represented in Fig. 2 by a synthetic spectrum with these parameters).  Then we see that, assuming log \gv=3.0, a good fit is obtained to TiO at 2900K, but to \na at 2550K.  On the other hand, assuming that log \gv=4.0 produces a good fit to TiO at 2750K but to \na at 3000K.  In the first case, \na implies a much lower temperature than TiO; in the second case, it implies a much higher one.  This is a direct consequence of the two facts discussed above:  \teff/gravity degeneracies act in opposite directions in TiO and \na, and \na is much more sensitive to gravity than TiO.  This allows rapid convergence onto the actual gravity.  In the example at hand, the disagreement between \teff(TiO) and \teff(\na) at both log \gv = 3.0 and 4.0, combined with the change in sign in [\teff(TiO)-\teff(\na)] in going from 3.0 to 4.0, would immediately lead us to try an intermediate gravity, log \gv=3.5.  At this gravity, we would infer 2800K from both the TiO and \na fits; log \gv=3.5, \teff=2800K would thus be the finally adopted, and correct, parameters.  

If our data were perfect, the complementary behaviour of TiO and \na discussed above would let us fix \teff and surface gravity to any desired precision.  However, real data do not allow this.  In particular, when comparing synthetic spectra to observations with finite resolution and non-zero noise, there is some intrinsic uncertainty in selecting the best-fit model.  We will illustrate these uncertainties in more detail later in this paper; for now, we simply state their magnitude.  At any given gravity, we cannot certify \teff to better than $\pm$ 50K from the TiO fits (e.g., in the example above, we would actually choose 2900$\pm$50 K at log \gv=3.0, 2800$\pm$50 K at 3.5, and 2750$\pm$50 K at 4.0, from the TiO fits).  Similarly, at a given \teff, we are insensitive to gravity variations smaller than $\pm$0.25 dex in the \na fits.  It is these uncertainties that finally set the precision of our derived parameters.  Note that these errors refer to the precision of our measurements, as distinct from the absolute {\it accuracy}, i.e. veracity, of the derived parameters; the latter is of course dependent on the validity of the synthetic spectra, which will be discussed in later sections.  

So far we have discussed \na; the situation is analogous for the \pot doublet (not shown).  In the latter case, a $\sim$ 100-150K decrease in \teff mimics a 0.5 dex increase in log \gv.  Ideally, \pot should be used, in conjuction with TiO, to provide an independent estimate of gravity which is as stringent as that from \na.  However, the \pot doublet lies significantly bluewards of \na, and thus has poorer signal-to-noise in some of our faintest and reddest objects.  In such cases, \pot is less effective than \na in constraining the gravity.  However, we emphasize that even in these cases, the gravity indicated by \pot is a good independent check on the gravity from \na; in all our targets, the two estimates are fully mutually consistent.

Finally, we also use the shape of the continuum surrounding the \na and \pot doublets to constrain \teff and gravity (this analysis is possible because our normalization procedure leaves the continuum shape intact).  As mentioned earlier, the continuum in M types is predominantly controlled by TiO; consequently, its behavior is very sensitive to \teff.  In particular, there are a number of temperature-sensitive TiO-bands immediately redward of both the \pot and \na doublets (e.g., $\lambda\lambda$7705, 7743 in the $\delta\nu$=-1 sequence just redward of the \pot doublet; $\lambda\lambda$8206, 8251 in the $\delta\nu$=-2 sequence just redward of the \na doublet; see \cite{Solf78}), which determine the continuum in these regions.  Fitting the continuum over the entire order that contains the \na or \pot doublet, therefore, permits us to break the degeneracy between \teff and gravity that affects the alkali doublets alone.  

This is illustrated in Fig. 2 for the \na order.  We see that, while the \na lines alone can be fit by various combinations of \teff and gravity, the surrounding continuum restricts the allowed range of \teff (and thus gravity).  In essence, by fitting an alkali doublet (line core and wings) together with a large swathe of surrounding continuum, we obtain a simultaneous fit to two different species - the alkali on the one hand and TiO on the other - each with different temperature and gravity sensitivities (TiO being largely dependent on \teff, and the alkali doublet depending on both \teff and gravity).  This greatly narrows the range of permissible \teff and gravity implied by the alkali fit.  In Fig. 2, we show that the continuum redwards of the \na doublet allows us to restrict \teff to at least $\pm$ 200K, and thus log \gv to $\pm$ 0.5 dex.  Unfortunately, the observed continuum in the \na and \pot orders contains a large number of closely-spaced telluric lines, which at times makes a determination of the stellar continuum harder.  For those of our objects most affected by telluric contamination, we cannot restrict \teff by this method to better than $\pm$ 200K; for the others, $\pm$ 150K can be achieved.  This is less accurate than the \teff constraints possible with the triple-headed TiO band.  Consequently, this band remains the best diagnostic for fixing \teff, and thus gravity when used in conjunction with the alkali doublets.  However, the continuua around the alkali doublets at least provide a very useful consistency check of the derived \teff and log \gv obtained from the triple-headed TiO band and alkali doublets, by restricting the allowed \teff to at least within $\pm$ 200K, and thus log \gv (from the alkali doublets) to $\pm$ 0.5 dex.

To summarize, we use the triple-headed TiO band and \na to simultaneously fix both \teff and gravity.  The \pot doublet provides an independent check of the derived gravity.  Fitting each alkali doublet together with the surrounding continuum additionally constrains the \teff and gravities, and provides a further check on the parameters derived from the TiO bandheads and alkali doublets.  Finally, we point out that our analysis also supports the validity of the synthetic spectra we use.  To derive \teff and surface gravity, we {\it demand} that the same model parameters fit all our spectral diagnostics.  It is not a priori clear that a given set of model spectra can actually satisfy this requisite - errors in the synthetic opacities may very well lead to inconsistencies between the fits to various diagnostics.  Thus, the fact that the model spectra used here yield consistent results - that is, for any given object, we are able to fit all our spectral diagnostics with the same \teff and gravity model (as we will show shortly) - suggests that our derived parameters are fairly accurate in an absolute sense.  

It is also very important to note that our \teff and gravity derivations are independent of extinction estimates.  Our high-resolution analysis involves comparisons between the observed and synthetic spectra over wavelength ranges of $\lesssim$ 100\AA.  Over any given spectral region this small, the wavelength-dependent variation in extinction is negligible for our purposes: the extinction over any such section can be assumed to be constant.  Of course, extinction will change the observed flux in our different spectral regions (TiO, \na, \pot), which are separated by more than 100\AA~ from each other, by different (multiplicative) constants.  However, our spectra are not flux-calibrated to begin with.  As described earlier, we have normalized the observed and synthetic spectra, in each of the spectral regions used in our analysis, by dividing by a numerical constant.  As a result, any extinction is effectively divided out of our data, and does not affect the final results of our analysis.  This is an important advantage of our method: when stellar parameters are derived by comparing the data to theoretical color-magnitude or HR diagrams, on the other hand, estimating the extinction is necessary a priori, and constitutes a significant source of uncertainty.  

\section{Results}
Our main results are: {\it (1)} the USco targets span a \teff range of $\sim$ 300K (2900--2600K); {\it (2)} the gravities for objects with \teff $\gtrsim$ 2750K are roughly similar, with log \gv $\approx$ 4--3.75 (commensurate with theoretical expectations for the age of USco, \S5.1); and {\it (3)} gravity declines sharply for cooler objects, with log \gv $\approx$ 3.5--3.25 (which is at odds with predictions for a coeval sample, \S 5.1).  We illustrate these results through a discussion of our model fits.   
     
In the interests of space, we will show only the best fits to a representative selection of objects from our sample.  However, to demonstrate our fitting procedure, and its robustness, we begin with a detailed analysis of two examples: USco 55 and USco 104 (\S 4.1).  In \S 4.1.1, we elucidate how their \teff and log \gv are obtained from fits to synthetic spectra, and discuss the precision of our measurements.  We show that our fits indicate a significant difference in gravity between the two.  To confirm the validity of this result in a model-independent fashion, we undertake direct comparisons between their observed spectra in \S 4.1.2.  In \S 4.2, we summarize the results for all our Upper Sco objects, and present more examples of gravity variations within our sample.  Various physical processes that could potentially skew our results are investigated in \S 4.3; we conclude that their effects on our analysis are unlikely to be significant.  The implications of our findings are discussed in \S 5; in particular, we examine the fits to GG Tau Ba and Bb in \S 5.4, in order to evaluate some of our hypotheses.  

\subsection{Two Test Cases: USco 55 \& USco 104}
\subsubsection{\teff and Gravity from Synthetic Spectra}
Consider first USco 55.  In Fig. 3 , we show fits at log \gv = 3.5 and 4.0 (4.5 is not shown, but discussed below).  At each gravity, we show the best fit model to each of our diagnostics.  All models have been rotationally broadened by a 12 \kms velocity profile, to match the observed \vsini of this object. 

At log \gv = 3.5, the best fit to the \na and \pot doublets is at \teff=2600K.  However, this is clearly not a good fit to the TiO bandheads, or to the continuum in the \na order.  Compared to the data, the triple-headed TiO band is much too strong in the model, especially at 8442 and 8452 \AA.  The TiO bands in the \na continuum are also too strong, depressing the model continuum relative to the data.  Conversely, a good fit to the TiO bandheads and \na continuum is found at this gravity with a \teff=2850 K model.  The \na and \pot doublets in this substantially hotter case, however, are much too weak and narrow to fit the observations.  The discrepancy between the \teff implied by the TiO bandheads and the alkali doublets is thus $\sim$ 250K, with TiO suggesting the hotter model.

A similar situation exists at log \gv = 4.5 (not shown).  The TiO bandheads, and the \na continuum, are now fit by a 2800 K model (note that this is very similar, but not identical, to the \teff required at log \gv = 3.5; this is due to the (very weak) \teff/gravity degeneracy exhibited by TiO, and alluded to earlier).   However, this model produces much stronger \na and \pot doublets than observed.  Instead, the best fit to the both alkali doublets is now found at $\sim$ 3000 K, which is a very poor fit to the TiO bandheads.  The mismatch between the \teff from the TiO and the alkali doublets at log \gv = 4.5 is thus $\sim$ 200K, about the same as at log \gv = 3.5.  However, unlike at log \gv = 3.5, TiO now indicates the cooler model.  This immediately suggests that the true value of gravity lies between 3.5 and 4.5.  Moreover, the TiO fits at both gravities indicate that \teff is somewhere in the 2800 to 2850K range.  

As illustrated in Fig. 3, these expectations are borne out using models with log \gv = 4.0.  At this gravity, good fits are obtained to all our diagnostic spectral regions in USco 55 -- TiO bandheads, \na and \pot doublets, and continuua surrounding the doublets -- at \teff=2800K.  We point out in passing the telluric lines slightly blueward of the blue lobe of \pot; there are also telluric lines blended into this lobe that make it appear very slightly broader than the model.  These lines are stronger in our fainter targets, and in general preclude our using the blue \pot lobe in the analysis in all but the highest S/N cases (such as USco 55).  At any rate, an excellent fit is obtained to at least the red lobe, as well as to the TiO and \na regions\footnote{Notice that at all the gravities and \teff shown, the models predict an absorption feature at $\sim$ 8196.5 \AA~, in the red wing of the \na doublet, which is absent in the data.  Given the enormous number of molecular lines in M-type spectra, and the as yet incomplete molecular data incorporated in the models, such minor discrepancies are not surprising.  As evident in Fig. 3, this particular discrepancy does not affect our \na fitting results, and we may safely ignore it.}.  Consequently, we adopt \teff = 2800K, log \gv = 4.0 for USco 55.

What are the errors in our derived parameters?  We address this in Fig. 4.  At our preferred gravity, log \gv = 4.0, we compare the observed TiO bandheads to a model at 2800K (our best-fit \teff), as well as to models offset from 2800K by 50 and 100K.  We see that a \teff variation of $\pm$ 100K makes a substantial difference: the 2700 and 2900K models are obviously bad fits to TiO.  Even a variation of 50K makes a (small) difference.  At 2850K, the model bandheads (particularly at 8442 \AA) are evidently stronger, and the continuum redward of 8452 \AA~ clearly depressed, compared to the data.  At 2750K, on the other hand, the model continuum is a good match to the data in the TiO order, but the 8452 \AA~ bandhead appears somewhat weaker than observed.  We therefore conservatively adopt a \teff error of $\pm$ 50K, at log \gv = 4.0, and rule out with high confidence deviations of $\pm$ 100K at this gravity.\footnote{A careful perusal of the 2800K and 2850K model fits reveals that a \teff of $\sim$2825K, instead of our adopted 2800K, may be marginally more appropriate:  the continuum redward of 8452 \AA~ is slightly better reproduced at 2850K, while the 8452 \AA~ bandhead is better matched at 2800K, suggesting an intermediate \teff.  However, this sort of $\sim$ 25K precision is only possible in our highest S/N targets (such as USco 55), and we do not pursue it further, except to note that \teff $\sim$ 2825K is within our estimated \teff range of 2800$ \pm$ 50K, and that a \teff closer to 2800 than 2850K is favored by our chi-squared tests.}.

Similarly, at our best-fit \teff of 2800K, we compare the \na doublet to a model at log \gv = 4.0 (our best-fit gravity), as well as to models offset from 4.0 by 0.25 and 0.5 dex (Fig. 4).  The models at log \gv = 4.0 $\pm$ 0.5 dex are clearly poor fits to the doublet.  Even a 0.25 dex change makes a small difference.  The model doublet at log \gv = 3.75 is (slightly) narrower than the data, while at log \gv = 4.25 it is clearly broader.  Given these results, we conservatively adopt a gravity error of $\pm$ 0.25 dex at \teff=2800K, and confidently rule out deviations of $\pm$ 0.5 dex at this \teff.  

The above error analysis does not explore the full parameter space:  we have examined TiO and \na separately, and investigated the effect on each of varying only one parameter (either \teff or log \gv) while keeping the other fixed at our best-fit value.  However, we have already shown previously (see Fig.3 and accompanying discussion in text), by varying both \teff and gravity and simultaneously comparing TiO and \na, that deviations of $\pm$ 0.5 dex from log \gv = 4.0 can be completely ruled out at any \teff.  Similarly, we can easily show that deviations of $\pm$ 100K from \teff = 2800K can be confidently ruled at any gravity.  To see this, imagine that \teff = 2900K.  The weak \teff/gravity degeneracy in TiO allows us to fit TiO at this temperature with log \gv = 3.25 (Fig. 4, bottom left, in blue).  However, the \na doublet can be be fit at \teff = 2900K only at a vastly different gravity of log \gv = 4.25 (Fig. 4, second panel from bottom right, in blue).  Similarly, if we assume a \teff of 2700K, a good fit to \na is obtained at log \gv = 3.75 (Fig. 4, second panel from top right, in blue).  However, no adequate fit can be found to TiO at this low \teff at any gravity; as discussed earlier, even log \gv = 4.5 requires a \teff of 2800K to fit the bandheads, while lower gravities require similar or higher \teff.  Thus \teff = 2800K $\pm$ 100K are completely untenable, regardless of gravity, by our requirement that TiO and \na give consistent parameters.  Similar tests (not shown), using TiO and \na inter-comparisons, indicate that changing both \teff and gravity by only 50K and 0.25 dex from the best-fit values give more consistent fits to the data compared to 100K or 0.5 dex variations, but still not as good as obtained with the best-fit model (the mismatches between models and data are similar to those shown in Fig. 4 for 50K and 0.25 dex variations).  All of this implies that our \teff and gravity are precise to at least $\pm$ 50K and $\pm$ 0.25 dex.  Finally, our chi-squared tests confirm our best-fit values of \teff=2800K and log \gv=4.0, as well as errors of at most 50K and 0.25 dex around these values.  We therefore adopt \teff = 2800K $\pm$ 50K and log \gv = 4.0 $\pm$ 0.25 dex for USco 55.

Now let us examine Usco 104.  In Fig. 5, we again show fits at log \gv = 3.5 and 4.0.  At log \gv = 4.0, the TiO bandheads are well matched by a 2750K model.  The \na doublet in this model, however, is too broad and deep to fit the data; the model \pot doublet is also slightly stronger than observed.  At this gravity, both alkali doublets are well matched by a 3000K model instead (strictly speaking, 3000K fits the \na doublet, while a slightly lower \teff, 2950K, fits \pot).  \teff $\sim$ 3000K, however, is obviously a poor fit to the TiO bandheads (Fig. 5).  Finally, the continuua in the \na and \pot orders are well fit by a 2750K model at log \gv = 4.0, in agreement with the \teff implied by the bandheads; 3000K is a poor fit to the continuua (this is shown in Fig. 5 for the \na order\footnote{The \pot continuum shown in Fig.5 does not clearly differentiate between the different \teff models.}).  The temperature discrepancy between the various diagnostics is thus $\sim$ 250K; log \gv = 4.0 is clearly not viable.  A similar analysis (not shown) also precludes log \gv = 3.0.  

At log \gv = 3.5, on the other hand, the TiO bandheads, \na and \pot doublets, and the continuua in the alkali orders are all fit very well by a 2750K model.  An analysis similar to that for USo 55 again yields \teff errors of $\sim$ $\pm$ 50K, and gravity errors of $\sim$ $\pm$ 0.25 dex.  For USco 104, therefore, we have \teff = 2750 $\pm$ 50K, and log \gv = 3.5 $\pm$ 0.25 dex.  These values are confirmed by our chi-squared tests.  

All the preceding analysis illustrates the efficacy of using inter-comparisons between the TiO bandheads and the \na doublet to determine temperature and gravity.  The strong \teff-dependence of TiO and the weakness of its dependence on gravity, combined with the strong gravity-sensitivity of \na and the opposite sense of the \teff/gravity degeneracy in TiO and \na (which causes the difference in \teff implied by the two diagnostics to change sign on either side of the most appropriate gravity), allows us to quickly zero in on the best \teff and gravity values.  Our fits also show the value of using the \pot doublet to check the gravity obtained from \na, and the usefulness of fitting the continuua surrounding the alkali doublets as a check on the \teff (and thus on the gravity).  With the best-fit model spectrum, all our spectral diagnostics - TiO bandheads, alkali doublets, and continuua - are fit remarkable well, while small deviations from the best-fit parameters cause large discrepancies between the models and data.  This attests to the robustness of our fitting procedure.  Finally, as mentioned earlier, the fact that a unique best-fit model can be found at all suggests that the synthetic spectra are quite accurate in absolute terms.  Errors in the model opacities can indeed lead to systematic offsets from the true \teff and gravity, when fitting any one of our diagnostic spectral regions.  However, any such offsets are more likely to be different for each of these regions; it is unlikely that the opacity errors all conspire together to produce a model spectrum that fits all all our diagnostics at the same \teff and gravity.  The presence of such consistency in our fits supports the accuracy of the derived temperatures and gravities.

\subsubsection{Empirical Evidence for Gravity Variation}
Notice, now, that we find USco 104 and USco 55 to have gravities differing by 0.5 dex.  We will soon see that even larger gravity differences also show up in our sample.  This is one of the main results of this paper, whose significance will be discussed shortly (\S 5).  For now, to drive home the point that this difference in gravities is real, and not an artifact of our fitting process, we directly compare the observed spectra of USco 55 and 104.  Such a comparison cannot, of course, yield absolute values for \teff and log \gv.  It can indicate, however, if the relative values we infer from model fits are reasonable.  In other words, we can test if the similarities and differences between the observed spectra are consistent with the model prediction of different gravities.  

In the top panel of Fig. 6, we compare the TiO bandheads in USco 55 and USco 104.  The continuua in the regions outside the bandheads (i.e., blueward of 8432 \AA) closely match each other, indicating that our relative normalization of the two spectra is good.  However, we see that the bandheads themselves, in particular at 8442 and 8452 \AA, are clearly somewhat stronger in 104, implying that its effective temperature is lower than that of USco 55.  This assertion is independent of any spectral-fitting, and is implied simply by the known behavior of TiO with temperature (the potential effects of dust are not considered here; we have noted before, and will argue in detail in \S 4.3.1, that grain formation is negligible in our targets).  We note that differences in rotational velocity cannot be responsible for the observed mismatch in bandhead strengths: model-independent cross-correlations show that USco 55 and 104 have very similar \vsini (12 and 16 \kms respectively; that their \vsini are indeed roughly equal is visually evident from the similar smoothness of the continuum molecular features in their spectra, in Fig. 6).  If anything, the slightly larger velocity of USco 104 should cause its TiO bandeads to be shallower, not deeper, than that of USco 55, if the intrinsic bandhead strengths were the same in both (in reality, tests show that the $\sim$ 4 \kms vsini difference should have an observationally negligible effect on the observed bandheads).  A difference in \teff is thus strongly suggested by the data alone.  This result qualitatively supports the (small, 50K) \teff difference we found from our detailed fits.   

Next we compare the \na doublets observed in USco 55 and USco 104 (Fig. 6, middle panel).    Clearly, \na in USco 104 is somewhat weaker (both narrower wings and shallower core) than in USco 55.  One might expect this if USco 104 were hotter:  increasing \teff would damp the abundance of both \na and the main perturber responsible for collisional broadening, \hydmol, producing a narrower, shallower doublet.  But the preceding TiO comparisons show that USco 104 is actually cooler than USco 55, which should make \na in 104 relatively stronger.  In other words, the \na doublet in USco 104 is comparatively weaker {\it in spite} of its lower \teff.  This strongly suggests that gravity in 104 is significantly lower:  by decreasing both the pressure-broadening by \hydmol, as well as the abundance of \na at any given optical depth, lower gravity leads to a weaker line; decreasing gravity sufficiently can compensate for the opposing effects of decreasing \teff on the \na doublet (as indeed illustrated by the synthetic spectra as well, in Fig.2; this is the source of the strong \teff/gravity degeneracy in \na in the models).  Once again, we point out that rotation effects cannot lead to the observed profile differences: first, because the two objects have almost identical \vsini, and tests show that the small \vsini difference would not have any discernible effect on these \na profiles; second, because the slightly faster rotation of 104, if observable, would actually lead to (very slightly) broader \na wings compared to USco 55, not manifestly narrower ones.  

We also note that the difference in \na profiles cannot be ascribed to normalization problems.  In the bottom panel of Fig. 6, we compare the entire \na order of USco 55 and 104.  The two continuua generally overlap very well, indicating good relative normalization.  Indeed, significant differences in the continuua occur mainly at, and slightly redward of, the TiO bandheads at 8206 and 8251 \AA; the bands are stronger in USco 104.  Thus, these TiO bandheads also imply that USco 104 is cooler than 55, just like the TiO examined above.

Taken together, therefore, the TiO and \na data alone strongly suggest that gravity in USco 104 is lower than in 55.  This result is an empirical one, independent of any spectral-fitting; it rules out the possibility that the difference in gravities derived from our detailed fits is only an artifact of our fitting procedure.  It is however conceivable that other real phenomena, e.g. cool spots, dust effects or varying metallicity, mimic gravity variations in the data.  Once we have examined the rest of our Upper Sco sample and presented more examples of apparent gravity differences, we will consider (and argue against) such effects.  

\subsection{Overall Results for Upper Sco}
The best fits to a representative selection of objects from our sample are shown in Fig. 7.  In each case, the `best fit model' is defined as one that simultaneously fits all our spectral diagnostics.  Specifically, we show the fit of this model to the triple-headed TiO bandhead and to the \na doublet (which together fix \teff and gravity), as well as to the \pot doublet and the continuum surrounding the \na doublet (both of which provide independent checks on the derived parameters).  We show only the red lobe of the \pot doublet, since in most of the objects shown, the blue lobe is degraded by telluric contamination.  A paucity of space also prevents our showing the fits to the continuum in the \pot order; in all cases, however, the same model is also the best fit to this continuum.  Our final adopted values are tabulated in Table 1.  We note that the models reproduce the data, especially the triple-headed TiO band, \na doublet and \na continuum, remarkably well.  This is a testament to the vast improvements in the atmospheric structure models, molecular and atomic linelists, and resulting synthetic spectra over the last several years.  

In the red lobe of the \pot doublet too, the fits are quite good, with two small caveats.  First, the data for USco 53 appears slightly broader than the model.  However, this is very likely due to an inadequately modeled absorption line at $\sim$ 7697 \AA~ (e.g., see best fit to USco 112 in Fig. 7, and USco 55 in Fig.3; the line appears well modeled in the former but less so in the latter), which is blended into the \pot red lobe at the high rotation velocity of USco 53, making the data seem a little broader.  Second, the fit to the \pot red lobe is clearly not as good in USco 109 and 130; this is due to the noisiness of the \pot data (we make this point explicitly by showing the unsmoothed data for these two object; the noise is obvious).  Both objects are at the faint and red end of our sample; as mentioned in \S 3.3, poorer S/N in the \pot order is thus to be expected, since it occurs significantly blueward of our other diagnostic regions.  Seeing conditions were moreover not optimal for USco 109; nevertheless, the best-fit model still reproduces the general shape of the line (except in the very core, where the data appears chopped by noise).  Even in USco 130, which is one of our two faintest and reddest objects, the model matches the data passably well (and reproduces the line core accurately).  For this object, we point out that the poorer quality of the \pot fit cannot be ascribed to a problem with the synthetic spectra at the relatively low \teff and gravity derived (from the TiO and \na fits): Fig. 11 shows an excellent fit to the \pot red lobe in GG Tau Bb, for which the S/N is much higher, at the same \teff and gravity; the poorer fit in USco 130 is indeed due to noise.  For both USco 109 and 130, the TiO and \na orders have much higher S/N, and the fits we obtain to these regions are clearly very good, and give consistent \teff and log \gv.  Moreover, we cannot improve the \pot fits in either object by choosing any other \teff/gravity combination; other such permutations also degrade the fits to TiO and \na.  We are therefore confident of the veracity of the \teff and gravity derived for USco 109 and 130 from the TiO bandhead and \na fits, in spite of the poor quality of the \pot data. 

Finally, we point out the good agreement between the models and data in three other \teff-sensitive TiO bandheads: at $\sim$ 8206 and 8251 \AA~ in the \na order, and at 7705 \AA~ in the \pot order (see Fig.7, and best-fits in Figs. 3 and 5).  At 8206 \AA~, the model bandhead seems slightly offset in wavelength from the data.  This is likely due to an inadequately modeled absorption feature just blueward of, and blended into, the model bandhead.  However, the overall strength of this bandhead is the same in the data and models.  At 8251 and 7705 \AA~ too, the agreement between the data and models is very good (the model-data agreement is perhaps best seen in USco 53, where the high \vsini smooths the stellar spectrum considerably, and allows the eye to avoid confusion from sharp noise and telluric features in the spectrum).  This shows the usefulness of using the continuum in the \na and \pot orders to constrain \teff, and supports the \teff (and thus gravities) we derive.  We note in passing that this is the first time such good agreement between synthetic spectra and observations has been demonstrated, over such relatively large swathes of continuum at {\it high}-resolution, in the M spectral classes; this is another sign of the accuracy of the models used here.  

Our model fits imply that \teff in our Upper Sco sample ranges from about 2900K to 2600K.  This is more or less consistent with expectations for young M5--M8 PMS objects (\cite{Luhman99}; we revisit this issue in later sections).  However, we also find that surface gravity varies from log \gv $\sim$ 4.0 to 3.25 (Table 1).  Six of our objects have log \gv $\sim$ 4, two have $\sim$ 3.75, one has $\sim$ 3.5, and two have $\sim$ 3.25 ($\pm$ 0.25 dex errors in all cases), with lowest gravity in the coolest targets.  This is seriously at odds with the predictions of theoretical evolutionary tracks, for a coeval sample of PMS objects at $\sim$ 5$\pm$2 Myr (the rough age of the Upper Sco cluster; \S 5.1).  While log \gv of 4.0--3.75 is commensurate with $\sim$ 5$\pm$2 Myr tracks, 3.5 or less is not, nor is a variation of $\sim$ 0.75 dex; we will discuss this shortly (\S 5).  First, we briefly discuss USco 104, 128 and 130, our coolest, lowest gravity objects.

For USco 104, we have presented a detailed spectral analysis in \S 4.1.2, in order to show that a lower gravity is commensurate with the data.  One can make a similar case for USco 128 and 130.  We explicitly discuss USco 130 here; the arguments are analogous for 128.  In Fig. 8, we compare the observed triple-headed TiO band and \na doublet in USco 130, to those in USco 55 and 104.  We note at the outset that rotational broadening effects are not germane to the following analysis: the difference in rotation velocity between USco 130 (\vsini $\approx$ 14 \kms) and either USco 55 or 104 (\vsini $\approx$ 12 and 16 \kms respectively) is $\sim$ 2\kms, negligible compared to other broadening effects (e.g., pressure broadening).  

First, we see that the TiO in USco 130 is significantly stronger than in 55 (especially in the $\lambda\lambda$ 8442, 8452 bandheads); this is why we have derived a lower \teff for it (2600K) than for 55 (2800K).  However, the strength of the \na doublet is very similar in both.  If both objects were the same gravity, then the \na doublet in the appreciably cooler USco 130 should instead be much {\it stronger} (i.e., broader and deeper).  The simplest explanation is that gravity is substantially lower in 130 than in 55, compensating for the much lower \teff of 130 and yielding very similar \na profiles for both.  Indeed, a closer look reveals that the doublet is slightly weaker (shallower core, narrower wings) in USco 130, suggesting that the temperature difference is more than compensated for by a gravity difference\footnote{This may not be clear in the plot shown, but is evident if the plot in the electronic copy is blown up.}.  This model-independent analysis is completely consistent with our fits to the synthetic spectra, which yield log \gv = 3.25 for Usco 130 and log \gv = 4.00 for USco 55.  

Similarly, we see that the TiO bands in USco 130 are clearly stronger than in 104, which agrees with our model fit result that the former is cooler than the latter (by $\sim$ 150K).  At the same time, we see that their \na doublet profiles are very similar.  Once more, if the two objects had the same gravity, \na in USco 130 should be much stronger, since it is significantly cooler.  This signifies that gravity in USco 130 is lower than in 104 as well, sufficiently so in order to offset the temperature differential between the two.  This is in agreement with our model fit results of log \gv = 3.25 in USco 130 and 3.50 in USco 104.  

In fact, a careful examination shows that the line-wings in USco 130 are slightly broader than in 104 (though by much less than might be expected from the large \teff differential between the two, were their gravities the same).  In other words, while the \na profiles in USco 130, 104 and 55 are all very similar, the doublet in 130 is more accurately intermediate in strength between that of 104 and 55 (which is why the difference between the USco 104 and 55 \na profiles is clearer, in Fig. 6, than the difference between the profiles of USco 130 and either 104 or 55, in Fig. 8).  What this implies is that, in the \na profiles, a large gravity difference between USco 130 and 55 over-compensates for the large \teff difference, while a nearly as large \teff differential between USco 130 and 104 is not completely compensated for by a smaller gravity differential; meanwhile, the effect of the much smaller temperature difference between USco 104 and 55 is overwhelmed by a large difference in gravity.  These conclusions, drawn from inter-comparisons of the TiO and \na data alone, are completely commensurate with the parameters derived from our fits: \teff=2800K, log \gv=4.0 for USco 55, \teff=2750K, log \gv=3.5 for USco 104, and \teff=2600K, log \gv=3.25 for USco 104.  The gravity variations in our sample implied by the synthetic spectra are thus not a spurious outcome of fitting uncertainties, but appear to be quite real.  

We point out in passing that gravity fluctuations may also affect precise spectral typing.  This is not very obvious in our sample, given the small range in spectral types covered (2.5 subclasses, derived by AMB00) and the relatively large typing uncertainties ($\pm$ 0.5 subclasses; AMB00).  Nevertheless, we note that USco 104 (M5) is one of the two earliest type objects in our sample, while we find it to be comparatively cool (2750K); this discrepancy may be due to its significantly lower gravity (log \gv = 3.50) compared to most of the other targets.  Our results suggest that gravity-induced typing uncertainties are not overwhelming: the two coolest objects (USco 128 and 130 at 2600K) are indeed assigned the latest types, despite having the lowest gravities.  However, we cannot rule out errors at the $\pm$1 subclass level, estimated from the range in types assigned by AMB00 to objects with different log \gv but same \teff (e.g., see targets at 2750K, Table 1).  This does not affect our analysis, since we do not base any conclusions on the precise spectral types (we only refer at times to the range in types covered, which is very likely to be in the right ballpark).  It prompts us to caution, nonetheless, that even in a roughly coeval sample, the ordering of objects by \teff and spectral type may not be identical, perhaps in part due to gravity effects, in the  cool young PMS regime; this is worth keeping in mind for PMS spectral type to \teff conversion schemes.  

Returning now to the veracity of gravity variations in our sample, we must pause to ask if any physical processes can mimic low gravity effects on the spectral line profiles, causing us to infer sytematically erroneous gravities (and \teff, for that matter).  We examine the four most plausible mechanisms: atmospheric dust, magnetically-induced cool spots on the stellar surface, metallicity variations, and changes in the pressure-broadening of alkali lines.  

\subsection{Physical Effects Mimicking Low Gravity}
\subsubsection{Dust}
In \S 3.1, we stated that the atmospheric models do not imply any dust formation in our PMS sample.  That is, the data are fit very well (as we have shown) by synthetic spectra in which grain effects are not yet important (at least in the red spectral regimes we have analyzed); dust appears in our model spectra only at still lower temperatures.  However, it is plausible that the treatment of grains in the models is inadequate.  Perhaps dust does matter in our coolest targets: the relative weakness of the alkali doublets in these objects is really due to overlying continuum grain opacity, and is misinterpreted as a signature of low-gravity by synthetic spectra that underestimate grain effects.  Is this suggestion viable?  We address this by comparing our mid- to late M PMS sample to field dwarfs of similar spectral type.  Dust opacity effects of the sort hypothesized here are known to be important in the latter (\cite{Jones97}; \cite{Leinert00}); an analysis of their spectra can thus indicate if grains contribute significantly to the spectral characteristics of our sample as well.

We perform this analysis in Appendix A, using a sample of field dwarfs that are homogeneous in gravity (log \gv $\sim$ 5.0), metallicity (solar) and \vsini, and span a similar range in spectral type as our PMS targets.  The main result is that we find no significant signatures of dust in the PMS objects analogous to those seen in the dwarfs.  Basically, the TiO bands in the dwarfs strengthen monotonically with decreasing \teff (i.e., later spectral type) until the onset of grain formation, at which point the bands reverse in strength and become monotonically weaker with later type.  Where the reversal occurs depends on the wavelength of the specific TiO bands considered: bluer bands (e.g., around 7050 \AA) are more strongly affected, and reverse strength at an earlier type, than redder bands (e.g., the $\sim$ 8440 \AA~ ones used in our preceding \teff analysis).   The strength of \na doublet in the dwarfs is seen to follow the behaviour of the 7050\AA~ TiO bands, reversing strength (though less markedly) as dust opacity kicks in.  Most of these effects have been pointed out by other investigators as well; the underlying reasons are briefly discussed in Appendix A.  In our PMS sample, however, we see no reversal in any the TiO bands with decreasing \teff, {\it in spite} of the weaker \na we have found in the coolest targets.  We conclude that dust cannot be responsible for the weaker alkali lines in the latter objects; gravity differences indeed seem to be the best explanation.  In fact, the 7050 \AA~ TiO bands (which are somewhat gravity sensitive as well, as spectral synthesis models indicate), while continuing to strengthen in our sample with decreasing \teff, without ever reversing, do exhibit additional variations that correlate well with the gravities we derive from \na; this further inplicates gravity differences within our PMS sample.  We mention in passing that a reduction in the neutral alkali strengths, such as we observe, cannot be ascribed to alkali depletion through molecule or grain formation: such effects, e.g., formation of alkali chlorides, appear only on progressing well into the L-types.

The absence of grain effects in our targets is not too surprising, for three reasons.  First, even if they were completely analogous to field M dwarfs, we would not expect to see very strong dust signatures even in our latest objects: at $\sim$ M7, they lie at the threshold of dust onset in field dwarfs.  Second, PMS objects of a given spectral type are thought to be somewhat hotter than dwarfs of the same type (e.g., \cite{Luhman99}); as we discuss later, the \teff derived in this paper are consistent with this view.  Higher \teff damps grain formation, so the onset of dust in the PMS phase might well occur at later spectral types than in field dwarfs.  Finally, our PMS objects have much lower gravity than field M dwarfs; the attendant decrease in photospheric pressure can inhibit the coalescence of molecules into grains (see \cite{Leinert00}).  At any rate, the absence of empirical dust signatures in our PMS spectra largely rules out grain effects skewing our gravity analysis.  

\subsubsection{Cool Spots}
We have undertaken a detailed analysis of the effect of cool spots on our \teff and gravity determinations; we summarize our results here.  We have considered changes in the observed line profiles induced by two phenomena: {\it (1)} the temperature contrast between a spot and the surrounding, hotter photosphere, and {\it (2)} the lower `effective' gravity of a spot compared to the true surface gravity, due to the added presence of magnetic pressure (and thus a decrease in gas pressure) within the spot.  We find that, even when spot parameters are chosen to maximize their effect on the observed optical line profiles - a spot areal coverage of 50\%, a temperature differential of 500K with the surrounding hotter photosphere, and a (resulting) effective gravity 0.75 dex lower than the true surface gravity - the effect on our inferred \teff and gravity is small.  Such spots cause us to derive a \teff lower by 150-200K than the photospheric temperature, and a gravity lower by 0.25 dex than the true surface gravity.  Whether the lower inferred \teff can be called erroneous is a matter of taste: when a cool spot covers half the surface, the total bolometric flux is indeed lower than that from an unspotted star; insofar as \teff is a measure of this flux, our derived lower-than-photospheric value is not inaccurate.  The difference between the true and inferred gravity is also not very high (the magnitude of this (systematic) change is of order the (random) error in our gravity determination).  Note that spots with either an appreciably larger or smaller temperature differential with the photosphere affect our results even less.

Secondly, even if we account for the possibility that our lowest gravities are actually underestimations by $\sim$ 0.25 dex, the gravity variations within our Upper Scorpius sample remain quite large ($\sim$ 0.5 dex).  Thirdly, our lowest gravity objects are also the coolest ones.  In \S 5.4, we will revisit why this might be; here, we point out that cool spots are not a viable explanation.  There is no reason to suppose that spots are preferentially present only on the low \teff objects, and not on the hotter ones as well (in fact, quite the contrary, if rotation affects activity, as usually supposed; see below).  Then, the gravities inferred in some of our hotter objects would also be too low, again by $\sim$ 0.25 dex (indeed, we do see variations at this level in hotter objects).  Spots may thus produce a spread of $\sim$ 0.25 dex in gravity over our sample, but not a (much larger) systematic difference in gravity between hotter and cooler objects.  All of this implies that invoking spots does not change our basic result, that large gravity differences appear within our Upper Sco sample with decreasing \teff.  Finally, available data suggests that any cool spots that are present should generally be smaller than we assume.  The covering fractions are usually of order a few percent; in the most spotted cases, the fraction may increase to $\sim$ 30\%.  Spot with covering fractions appreciably smaller than the 50\% we assume would affect our results negligibly. 

Finally, the covering fraction of cool spots is expected to increase with rotational velocity.  However, USco 104, 128 and 130, the three objects with anomalous gravity (compared to the theoretical tracks) are not especially fast rotators.  Indeed, USco 128 is the {\it slowest} rotator in our sample, while USco 104 and 130 have moderate velocities, comparable to those in other Upper Sco objects that do not exhibit any gravity anomalies.  It is true that we only have \vsini; the true equatorial velocities of USco 104, 128 and 130 may be higher.  By the same token, however, we would expect to see gravity variations among those objects in our sample with the highest \vsini.  This is not observed: though some of our hotter targets have \vsini approaching, or even exceeding 50 \kms, none of them evinces gravity anomalies.  In short, if large cool spots are to blame for the behaviour of USco 104, 128 and 130, one would have to postulate that such spots only besiege cool, slow to moderate rotators, and {\it not} slightly hotter, much faster rotating ones.  This does not appear very likely: on the one hand, there is no particular reason to suspect such a strong \teff-dependency in spots; on the other hand, there are good physical reasons to expect {\it faster} rotators to be more spotted.

In summary, cool spots are very unlikely to affect our \teff and gravity derivations.  Even if they result in some small spread in gravity over our sample, they are unlikely to cause the large drop in gravity we find in the coolest objects within the sample.  

\subsubsection{Metallicity}
In this work, we have assumed solar-metallicity (i.e, used synthetic spectra with [M/H] $\equiv$ $log\left[\frac{{\rm{M}}/{\rm{H}}}{{\rm{M}}{_{\odot}}/{\rm{H}}{_{\odot}}}\right]$ = 0.0).  However, variations in metallicity can potentially affect the temperatures and gravities we infer.  Specifically, higher metallicity reduces the abundance of hydrogen particles (which are the main source of collisional broadening); it also implies a decrease in pressure at a given optical depth (because of higher opacity).  Both effects tend to produce a narrower alkali line, just as {\it decreasing} gravity does (\cite{Schweitzer96}; \cite{Basrimoh00}).  Simultaneously, higher metallicity can also mimic lower \teff, by making the TiO bands stronger (\cite{Leggett92}), both due to an increase in abundance of Titanium and Oxygen, and a decrease in temperature at a given optical depth (again, due to higher opacity).  In summary, not accounting for an increase in metallicity can yield spuriously low \teff and gravity.  Can this effect explain the presence of low-gravity objects at the cool end of our sample?  We do not think so, for the following reasons.

Using synthetic spectra to fit the observed \na and \pot optical doublets (i.e., the same technique employed in this paper), \cite{Schweitzer96} have examined in detail the effect of varying metallicity on the gravity inferred for the M8 field dwarf VB 10.  They show that for fixed \teff (2700K in their analysis, similar to the \teff derived for our coolest objects),  increasing (decreasing) metallicity by 0.5 dex has the same effect on the alkali lines as decreasing (increasing) gravity by an equal amount. The implications of this are two-fold:  {\it (1)} a change in metallicity is certainly capable of masquerading as a change in gravity; however, {\it (2)} the magnitude of the required metallicity shift is comparable to that of the perceived shift in gravity.  Now, we derive log \gv $\approx$ 3.25$\pm$0.25 dex for the lowest gravity objects (USco 128, 130), while the rest of the sample is mostly at $\sim$ 3.75--4.00; the latter is also the range predicted by the theoretical tracks for mid- to late M PMS objects at an age $\sim$ 5$\pm$2 Myr (\S 5.1).  Invoking metallicity variations instead then implies USco 128 and 130 are over-abundant, compared to both solar and the rest of our Upper Sco sample, by $\sim$ 0.75 dex, and by at least $\sim$ 0.25 dex even taking the lower limit on predicted gravity and our $\pm$0.25 dex errors in log \gv into account.

This is quite a substantial variation in abundance.  Such a deviation from solar metallicity would be surprising for a young, nearby association; moreover, differences at this level are even more unlikely {\it within} a single association.  Indeed, \cite{Padgett96} have analysed a number of nearby star-forming regions (Taurus-Auriga, Ophiuchus, Chameleon, Orion), and found solar abundance to within $\lesssim$ $\pm$ 0.1 dex (i.e., $\pm$ 25\%) in all of them; within a given region, the variation is at most at the same level.  In light of this, the over-abundance required in USco 128 and 130 ($\sim$ 2--4 times solar) to produce the gravity-effect we see appears implausibly high.  Finally, we will show further on (\S 5.4) that a similar deviation from the tracks is found in the GG Tau B system as well: while gravity in GG Tau Ba agrees quite well with the predicted value, Bb has a gravity lower than expected by $\sim$ 0.25--0.5 dex.  Since Ba and Bb not only belong to the same star-forming region (Taurus) but also to the same bound stellar system, this effect is highly unlikely to result from a large difference in metal abundance between the two stars.  These arguments lead us to conclude that metallicity variations are not a viable explanation for the range in gravities, and consequent deviation from the theoretical evolutionary tracks (see \S 5.1), implied by our spectral analysis.  

\subsubsection{Collisional Broadening}
Through inter-comparisons of the data alone, we have shown that the coolest objects in our sample have weaker alkali lines than hotter ones.  We have argued that this cannot be due to the onset of dust opacity.  The only other way to achieve this effect is through a weakening of the line broadening mechanism in our coolest targets.  This can happen in three ways: {\it (1)} if the intrinsic gravity in the coolest objects is lower, or {\it (2)} their alkali lines arise mainly in cool spots with lower gas pressure than the surrounding photosphere, or {\it (3)} their metallicity is higher.  Comparing our data to the synthetic spectra, we find that the effective reduction in gravity required to reproduce the alkali profiles in our coolest targets is of order 0.50--0.75 dex:  either the gravity of these objects is truly lower by this amount than in the rest of the sample, or cool spots or metallicity variations mimic a gravity reduction of this magnitude.  We have argued that spot or metallicity effects are highly unlikely to produce such a large offset, implying a real variation in intrinsic gravity.

However, one might argue that, while the empirical data comparisons certainly indicate a reduction in line broadening at the cool end of our sample, the {\it magnitude} of this effect is overestimated by the synthetic spectra.  As noted earlier, metallicity variations (and thus inferred gravity fluctuations) of $\sim$ 0.1 dex cannot be ruled out; cool spots may also mimic gravity effects at the $\lesssim$ 0.25 dex level.  Moreover, a reasonable age spread of $\sim$ 1--2 Myr in the Upper Sco sample can lead to real shifts in gravity of $\lesssim$ 0.25 dex (see \S5.1 and Fig. 9).  Is it possible, therefore, that the gravity in the coolest objects is effectively only $\sim$ 0.1--0.25 dex lower than in the hotter objects, and that the model spectra suggest a much larger variation due to an incorrect treatment of line broadening?  We think not, for the following reasons.

Spots, metallicity differences and/or an age spread should affect our entire USco sample equally.  However, while a 0.25 dex spread in derived log \gv is indeed seen in our hotter targets (\teff $\gtrsim$ 2750K), and may arise from any or all of these causes, we do not find larger, dramatic departures from expected gravities in this \teff regime - as we show in \S 5, our log \gv for these objects are all in agreement with the evolutionary track predictions for the assumed age of Upper Sco (5$\pm$2 Myr).  This implies that the model treatment of line broadening at these temperatures is adequate; to produce the much larger drop in log \gv derived for our coolest objects, the synthetic spectra must grossly mistreat broadening only at the lowest \teff encountered here.  Our entire sample, though, covers only 2.5 spectral subclasses ($\sim$ M5--7.5), signifying a quite small range in \teff.  Correspondingly, we find a temperature spread of $\sim$ 300K; regardless of the absolute accuracy of our derived \teff, this is commensurate with the range estimated by other investigators for these spectral types, both at PMS ages (\cite{Luhman99}) and in the field (e.g., \cite{Leggett01}).  Thus, significant errors must appear in the synthetic treatment of line-broadening over a \teff change of only $\lesssim$ 300K.

This does not seem viable.  The neutral alkali lines in M-type spectra are predominantly affected by van der Waals (vdW) collisional pressure broadening, with the main vdW perturbers being \hydmol, \hyd and \hel (\cite{Schweitzer96}).  Of these, \hydmol is by far the dominant contributor in M-types; nevertheless, all three are treated in the model spectra.  The broadening due to \hydmol, and the other perturbers, varies smoothly with temperature, and is not expected to change drastically over the narrow \teff range covered by our sample.  Thus, any broadening-related systematic errors in our log \gv are likely to be nearly constant over the sample.  In other words, if the agreement between our results and the evolutionary model predictions for objects with \teff $\gtrsim$ 2750K is valid, it is difficult to imagine, {\it without} doing violence to these results, that the line-broadening treatment falters significantly for the 2750--2600K targets in which we find low gravities.  An inadequate modeling of line-broadening is unlikely to yield spuriously low gravities only at the cool end of our sample.

\section{Implications}
In \S 5.1, we compare our derived \teff and gravities to the predictions of theoretical evolutionary tracks constructed by the Lyon group, which are currently the most widely used for low mass stellar and substellar studies.  We find significant discrepancies, implying either {\it (1)} problems in our analysis, or {\it (2)} a substantial \teff-dependent age spread in our sample, or {\it (3)} problems with the theoretical tracks.  Our arguments in \S 4.3 indicate that, while there is certainly room for errors in our analysis, these do not seem likely to cause the large divergence between our results and the evolutionary predictions.  We therefore concentrate on the two other possibilities.  In \S 5.2, we discuss various scenarios whereby a real, large age spread might arise; none seems particularly convincing.  We further undermine the hypothesis of a real age spread by demonstrating, in \S 5.3, that similar age discrepancies also appear for GG Tau Ba and Bb, which are almost certainly coeval.  In \S 5.4, we consider the remaining possibility - uncertainties in the evolutionary models - by examining some likely problems in the choice of initial conditions and treatment of convection.  

\subsection{Comparison to Theoretical Evolutionary Tracks}
Recently, Preibisch et al. (2002) have derived a well-constrained age of 5$\pm$2 Myr for the Upper Scorpius association.  Their sample consists primarily of G, K and early M stars (i.e., somewhat higher mass objects than ours); the age has been derived through comparison with the theoretical isochrones of \cite{Dantona94} and \cite{Baraffe98}.  The Preibisch et al. result is fully consistent with the 5-6 Myr derived from the Main Sequence turnoff of A and B stars in Upper Sco (\cite{dezeeuw85}; \cite{degeus89}).  It also agrees with the kinematic age of $\sim$ 5 Myr, derived by tracing back the motions of stars in the association till the smallest configuration is reached (\cite{Blaauw91}).  

With these results in mind, we compare our derived \teff and gravities to those predicted by the theoretical evolutionary tracks of the Lyon group.  In particular, we combine the models presented in \cite{Baraffe98} (hereafter BCAH98) and \cite{Chabrier00a} (hereafter CBAH00).  The BCAH98 models cover masses from 0.02 to 1.4 \msun, while the CBAH00 ones extend from 0.001 to 0.1 \msun.  Moreover, the latter models use synthetic spectra (as an outer boundary condition) that incorporate dust formation, while the spectra used in the older BCAH98 models neglect grains altogether\footnote{We note that the synthetic spectra adopted in the BCAH98 and CBAH00 tracks, including the `Dusty' spectra, are significantly older generations than the ones considered in this paper, with some substantial changes in model opacities in the intervening years}.  However, for the range of masses over which the two sets of tracks overlap, 0.02--0.1 \msun, and at least over the 1--10 Myr age range of interest in this paper, the temperatures and gravities predicted by both are almost identical, i.e., dust does not affect the predictions of these tracks at least above 0.02 \msun for these ages.  In order to compare our results to the theoretical predictions for a large range of masses, therefore, we have combined the BCAH98 and CBAH00 tracks.  Specificlly, we have adopted the BCAH98 models for masses greater than 0.02 \msun, and the CBAH00 models for lower masses.  For the sake of concision, we will henceforth refer to this merged set of tracks as the Lyon98/00 model.  One other point needs to be made, regarding the treatment of convection in the evolutionary tracks.  BCAH98 models come in 3 mixing-length flavors: $\alpha$ = 1.0, 1.5 and 1.9.  However, only the $\alpha$=1 models extend to masses below 0.6 \msun.  Similarly, the CBAH00 tracks, which are restricted to masses $\leq$ 0.1 \msun, exclusively use $\alpha$=1.  In other words, evolutionary models with $\alpha$$\neq$1 are {\it not} available for masses $<$ 0.6 \msun, which is precisely the mass-regime of interest for our mid- to late M PMS sample.  Consequently, the Lyon98/00 merged tracks we show are only for $\alpha$=1.  This will be important further on, when we attempt to understand discrepancies between our results and the model predictions in terms of uncertainties in the model treatment of convection.    

The comparison between our results and the tracks is shown in Fig. 9.  Three facts are immediately evident.  First, in objects hotter than $\sim$ 2750K, our gravities (log \gv $\approx$ 3.75--4.00) are commensurate with those predicted by the 5$\pm$2 Myr tracks (indeed, consistent with precisely 5 Myrs, within our $\pm$0.25 dex measurement uncertainty). Second, we find a sharp drop in gravity for objects cooler than $\sim$ 2750K; USco 104, 128 and 130 are all much lower in log \gv than hotter targets.  Third, no such drop is predicted by the theoretical tracks, for roughly coeval objects at 5$\pm$2 Myr.  The Lyon98/00 models imply that, for objects with \teff $\approx$ 2200--3200K, gravity variations arise predominantly from differences in age: 5$\pm$2 Myr-old bodies at these temperatures should all have log \gv $\sim$ 3.75--4.00.  Consequently, when combined with our derived parameters, the tracks predict a large age discrepancy between the hotter and cooler targets.  Objects cooler than $\sim$ 2750K appear substantially younger than 5$\pm$2 Myr due to their lower gravity.  USco 104 (log \gv $\approx$ 3.5) seems to have an age $\lesssim$ 1 Myr, while USco 128 and 130 (log \gv $\approx$ 3.25) appear even younger.  We emphasize that sliding our lowest gravity objects along the \teff axis - i.e., assuming a systematic error in our derived \teff - leaves the discrepancy with the tracks largely unchanged: any given isochrone in the 5$\pm$2 Myr range changes by $<$ 0.25 dex in log \gv over the entire 1000K range shown.  Our divergence from the tracks implies a serious disagreement with the predicted gravities.

As an aside, we note that the evolutionary tracks predict masses corresponding to the \teff and gravities we find; the tracks for various masses are plotted in Fig. 9.  However, these are not necessarily the masses we derive eventually (Paper II), from a consideration of the observed photometry of our sample combined with our gravity estimates.  In general, we will find that our USco masses are in the same ballpark as the model predictions (within the factor of 2 uncertainty in our mass determinations; Paper II).  However, for USco 128 and 130 we will derive masses of $\sim$ 9--14 \mj (within a factor of 2), substantially lower than the model prediction of $\sim$ 30 \mj (if the mass tracks in Fig. 9 are extrapolated down, at constant \teff, to our gravity for these two objects). For GG Tau Ba and Bb, which are also plotted in Fig. 9 and will be discussed shortly in this paper (\S 5.3), the models predict (again, extrapolating at constant \teff in Fig. 9) masses of 0.04 \msun (Ba) and 0.02 \msun (Bb); the final mass we derive in Paper II for Ba is significantly higher in comparison (0.12 \msun), but quite similar for Bb (0.028 \msun).  We will revisit this issue in detail in Paper II; for now we simply caution that the mass tracks shown in Fig. 9 are for comparison only, and do not signify the final masses we adopt for our sample.  The real value of the \teff / log[g] plot in Fig. 9 lies in revealing that our gravities for GG  Tau Bb, USco 128 and 130 (and to a lesser extent, USco 104) are completely inconsistent with the model gravities, for the estimated ages of the Taurus and Upper Sco complexes, {\it regardless} of the precise mass (or \teff) we assign these objects over the entire 0.2--0.01 \msun mass range shown (as well as at still lower masses; not shown here but see Paper II). 

There are three possible explanations for the age (or equivalently, gravity) discrepancy between our results and the theoretical predictions.  {\it (1)} Our spectral analysis is incorrect; {\it (2)} there is a real age-spread in our Upper Sco sample, with cooler objects being younger; or {\it (3)} the evolutionary tracks are problematic.  However, we have demonstrated in \S 4, through inter-comparisons of the data alone, that our coolest USco objects indeed exhibit spectral signatures of relatively low gravity.  We have further argued that these signatures are unlikely to arise from dust opacity, cool spot or metallicity effects.  Finally, through a consideration of pressure-broadening in the alkali lines, we have argued that if our gravities for the hotter objects are correct (which allows them to agree with the Lyon tracks), then the magnitude of the gravity drop we derive for the coolest targets is likely to be roughly accurate as well.  We therefore proceed under the assumption that our spectral analysis is by and large correct, and turn our attention to the second and third of the possible explanations listed above.  

\subsection{Scenarios for Real Age Variations}
If USco 104, 128 and 130  are truly $\lesssim$ 1 Myr old, then they are either not members of the Upper Sco association, or some star formation has occurred much more recently in Upper Sco than previously suspected.  Now, all the evidence firmly points to USco 104, 128 and 130 being PMS objects (see \S 2.1).  In that case, if they are not members of the association, they must either be PMS objects in the foreground or background that appear projected onto Upper Sco, or they must have physically travelled into Upper Sco from some other nearby star-forming region.  The projection scenario is unlikely: there is no obvious evidence, from all the studies of PMS objects in Upper Sco so far, of contamination by background or foreground {\it PMS} objects.  It would be very strange then, if in a small sample of 11 we found 3 such objects.  On the other hand, the $\rho$ Ophiuchus dark cloud borders Upper Sco to the east, and is known to harbor PMS objects younger than a million years.  Is it possible that USco 104, 128 and 130 are actually members of $\rho$ Oph?  We cannot discount this possibility without proper motion data. Moreover, in the few $\rho$ Oph T Tauris with measured radial velocities, the value ($\sim$ -5 \kms; G. Doppmann, 2002, private comm.) is similar to that in Upper Sco, so our \vrad measurements cannot shed any light on the issue.  

However, we can make the following plausibility argument.  The core of the $\rho$ Oph dark cloud lies $\sim$ 25', and the western edge $\sim$ 20', from the observed positions of our three objects.  Thus, if they are really $\lesssim$ 1 Myr old, and originally members of $\rho$ Oph, they must have space motions in excess of about 25 \kms, in order to have reached their present locations in the time since their birth.  This is an implausibly large value.  Given that the velocity dispersion of the molecular gas in $\rho$ Oph is a few \kms, the chance of any single isolated object, let alone three, having such a large velocity is miniscule.  Indeed, 25 \kms is of order the escape velocity from within an AU of a solar mass star.  Ejection, through dynamical encounters, is thus a possibility.  However, we consider it highly improbable that 3 out of our 11 Upper Sco objects happen to be very young $\rho$ Oph members, originally in close orbit around a solar mass star, and eventually flung out in the direction of Upper Sco.  Each of these steps has a very small probability, and their combined probability appears negligible. Proper motions are needed to settle this issue definitively.  

The other way of accomodating the apparent youth of USco 104, 128 and 130 is to postulate very recent star formation in Upper Sco.  While we cannot rule this out, it seems very unlikely, given that there are no signs of this in the \cite{Preb02} study.  In other words, if 3 out of our 11 objects, or $\sim$ 25\% of our sample, are so young, we would expect this effect to show up strongly in the latter study's much larger sample.  It does not, enabling \cite{Preb02} to strongly constrain the age of the association to 5$\pm$2 Myr.  On the other hand, it is possible that real age variations are preferentially present at the latest spectral types; since the Preibisch et al. study consists predominantly of earlier spectral types than ours, this would explain why we see an effect while they do not.  Is this likely?  If one assumes that spectral type corresponds roughly to mass\footnote{Notwithstanding the spectral typing uncertainties we have discussed earlier, there is not much doubt that the mid- to late M stars in our sample are on average cooler than the early M (and earlier) stars in the Preibisch study.  In the early contraction phase, evolutionary models indicate (see Fig. 9) that objects of a given (low) mass descend roughly along Hayashi tracks, with \teff approximately constant or, as one moves deeper into the substellar regime, decreasing with time.  At a given age, cooler objects are then less massive than hotter ones, and {\it younger} cooler objects less massive still.}, then a spectral type-dependent star-formation history implies a mass-dependent one.  It is hard to imagine a scenario, however, that leads to the sudden birth of only very low-mass isolated objects, 5 Myr after higher mass stars have ceased to form.  The Upper Sco cluster is also relatively free of molecular gas, further undermining the possibility of recent star formation.  

On the other hand, it may be that USco 104, 128 and 130 have not formed in isolation at all, but rather in accretion disks around some other Upper Sco members within the last million years, and subsequently been ejected.  This would explain their relative youth, as well as remove the necessity of postulating recent isolated star formation in Upper Sco.  There are indeed other Upper Sco PMS objects in the vicinity of USco 104, 128 and 130, around which the latter three may originally have formed.  However, formation in an accretion disk, allied with an age much less than the parent star, seems plausible only if these three objects have roughly planetary masses.  We will return to this scenario in \S 5.4.   

In summary, we see that a real age spread in our Upper Sco sample appears largely unlikely, but cannot be completely ruled out.  To test whether such a spread is indeed responsible for the discrepancy between our results and the theoretical tracks, it is better to consider objects which are almost certainly strictly coeval, such as different components of a single stellar system.  We thus proceed to analyze GG Tau Ba and Bb, the two M-type members of the quadruple GG Tau system in Taurus.  Our reason for choosing this specific system is that it has been studied extensively by other investigators; in particular, \cite{White99} (hereafter, WGRS99) have previously compared its components to various evolutionary tracks, providing a frame of reference for our present analysis.  At any rate, if our results in this case once again differ from the theoretical predictions, mirroring the situation in Upper Sco, then a substantial age spread specific to Upper Sco cannot be readily invoked to resolve the diagreement, and an explanation must be sought elsewhere.   

\subsection{\teff and Gravity in GG Tau B: Implications for Age Discrepancies}
Our method of \teff and gravity determination for GG Tau Ba and Bb is analogous to that adopted for Upper Scorpius.  The complete analysis is presented in Appendix B, and the fits obtained, which are excellent, shown in Fig. 11 ; here we only state the results.  First, we find \teff for Ba and Bb to be $\sim$ 2800K and 2600K respectively\footnote{More precisely, we find 2775$\pm$50 K for Ba and 2575$\pm$50 for Bb.}; these are $\sim$ 200K lower than the previous estimates by WGRS99, acquired through comparison with the BCAH98 tracks.  Second, while our gravity for Ba (log \gv $\approx$ 3.4), together with our derived \teff, gives a plausible age of $\sim$ 1 Myr for this object (within our $\pm$ 0.25 dex uncertainty in log \gv), our gravity for Bb (log \gv $\approx$ 3.1) makes it seem much younger.  Finally, even adopting the higher WGRS99 temperature for Bb does not bring it into agreement with the 1 Myr track (though it improves the situation somewhat).  These findings are depicted in Fig. 9.  Our GG Tau result, therefore, is completely analogous to our Upper Sco one:  there is a drop in gravity as we move from an object hotter than $\sim$ 2750K to one that is significantly cooler, and this drop, not predicted by the theoretical evolutionary models for coeval bodies, causes the models to imply a much lower age for the cooler object. 

We pause here a moment, to contrast our results with those of WGRS99 (see Appendix B for details).  The latter authors plot GG Tau Ba and Bb on the theoretical Hertsprung-Russell diagram (luminosity versus \teff) predicted by the BCAH98 models, and state that these models are consistent with coevality in the GG Tau system.  At first glance, this seems to directly contradict our result that the BCAH98 (or equivalently, Lyon98/00) tracks are {\it not} commensurate with coevality.  However, the WGRS99 study is not, strictly speaking, a test of the tracks, since they do not fix the \teff for Ba and Bb a priori.  Instead, the temperatures are only constrained to a 400K range, defined by the \teff in giants and dwarfs of the same spectral type as Ba and Bb.   Hence, what WGRS99 really find is the following:  {\it assuming} coevality, the Lyon98/00 tracks predict \teff for GG Tau Ba and Bb that appear reasonable, i.e., lie somewhere between the dwarf and giant temperature scales.  Indeed, \cite{Luhman99} carries out an identical analysis (assume coevality in GG Tau, in conjunction with the Lyon98/00 tracks) to {\it define} a PMS \teff scale.  Without better a priori constraints on the \teff of Ba and Bb, WGRS99 do not strictly test whether the Lyon98/00 tracks actually predict coevality.

Our analysis is analogous to that of WGRS99, albeit with gravity replacing luminosity as one of the two parameters in the HR diagram.  The difference is that our \teff and gravity for Ba and Bb are fixed in advance, independent of the Lyon98/00 models.  Consequently, ours is stringent test of the evolutionary tracks, to the extent that the \teff and log \gv we infer from spectral fitting are accurate.   We find that, when confronted with our more precisely determined parameters for the two objects, the Lyon98/00 tracks no longer predict coevality.

Can Ba and Bb really have different ages?  Most of the formation scenarios we considered earlier for Upper Sco (e.g., staggered star-formation), that might lead to such age variations, are not applicable to GG Tau Ba and Bb, since they clearly belong to the same stellar system.  However, we also raised the possibility that planemos formed later in an accretion disk may be much younger than the parent star.  Can this explain the much lower age inferred for Bb from the Lyon98/00 tracks?  It appears not; in Paper II, we derive a mass of $\sim$ 0.03 \msun for Bb (independent of theoretical tracks), comfortably higher than a planemo. So far, nobody has suggested that brown dwarfs in binaries form in accretion disks much later than the parent star: even if they form from accretion disk instabilities (\cite{Bate02}), the process probably occurs very early in the system's life - both because that's when the disk is massive enough to be (gravitationally) unstable, and also to ensure that there's enough unaccreted disk material left to form the brown dwarf.  Thus, large age differences between the two components of GG Tau B seem unlikely to arise from differences in formation history.  

The fact that we find the same disagreement with the tracks in Upper Sco as well as in GG Tau - the coolest objects appear implausibly young - suggests that the same underlying mechanism is responsible in both.  That is, we may reasonably rule out a real age spread that would affect Upper Sco alone (e.g., young interlopers from $\rho$ Oph).  The one remaining explanation is that the theoretical evolutionary tracks are somehow inaccurate for these very young low-mass objects.  This would not be too surprising, as has been pointed out by the theoretical modelers themselves \citep{Baraffe02}: the lower in mass and age one goes, the more suspect are the calculations (especially for objects which have essentially just formed).  We thus turn to a discussion of evolutionary model uncertainties, and whether they can credibly explain our results.   

\subsection{Evolutionary Models Uncertainties}
\cite{Baraffe02} have recently discussed various sources of uncertainty in current theoretical evolutionary models for very young, low-mass PMS objects.  The three most eggregious ones they identify are the choice of initial conditions, accretion effects, and the treatment of interior convection.  To this we will add uncertainties in deuterium-burning conditions.  Let us examine initial conditions first.

In current evolutionary models, the calculations are begun at (i.e., a time $t$=0 is assigned to) some specified point during the initial, fully convective, quasi-hydrostatic contraction phase of low mass stars and substellar objects.  The choice of this starting point is at present somewhat arbitrary; however, it obviously affects the model age (defined as time elapsed since $t$=0) implied for an object with an observed set of physical conditions: if the calculations are begun earlier in the contraction phase, the implied age will be larger.  This uncertainty is nicely illustrated by \citet{Baraffe02} through a comparison of models with varying initial gravities (for a given mass).  Can this effect be responsible for our results?  

The initial conditions assumed in the Lyon98/00 models are as follow.  For objects with mass $\gtrsim$ 0.03 \msun, the calculations begin at the point where deuterium (D) fusion starts; for lower masses, they begin at log \gv = 3.5 (Baraffe 2003, pvt. comm.; see also \citet{Baraffe02}).  Note in this context that, in the Lyon models, D-fusion begins at log \gv $\sim$3.5 in a 0.1 \msun object, at log \gv $\sim$3.75 in a 0.03 \msun object, and at log \gv $\sim$4.0 in a 0.02 \msun object.  Moreover, in this mass range, D-burning lasts from roughly 1--15 Myr, with the fusion timescale increasing with lower mass \citep{Chabrier00b}.  During fusion, \teff and luminosity remain practically constant (i.e, this is the `Deuterium Main Sequence'); hydrostatic contraction resumes once the deuterium is exhausted.  Objects less massive than about 0.012 \msun (12 \mj), on the other hand, are below the minimum D-burning limit: they never achieve D-fusion or halt in their contraction (until core degeneracy sets in much later).  Given these considerations, we can estimate the errors in age in the Lyon models, as follows.  

The Kelvin-Helmholtz contraction timescale (\tkh), applicable to our objects before D-fusion begins, is [3G{{M$_{\ast}$}$^2$}/7R$_{\ast}$]/[4$\pi${{R$_{\ast}$}$^2$}$\sigma${\teff$^4$}] (valid for fully convective objects).  In our case, where gravity and \teff are the observables, and the masses, gravities and \teff are low, it is profitable to express it as:

$$ t_{KH} \approx 2 \,\,{ {\left[{\frac{g}{10^{3.5} \,{\rm{cm/s^{2}}}}}\right]^{3/2}} {\left[{\frac{M_{\ast}}{0.03 \,{\rm{M}}{_{\odot}}}}\right]^{1/2}} }{\left[{\frac{{\rm{T}}_{eff}}{2700 \,{\rm{K}}}}\right]^{-4}} \, {\rm{Myr}} \eqno{1} $$

Now, eqn. [1] is only applicable at a given instant in time; to make use of it, we need to connect it to the real age ($\tau$) of an object, where $\tau$ is defined as the time it takes to contract from an infinite (i.e., relatively very large) radius to its present size.  This is easily done if contraction (at any given mass) can be assumed to proceed at roughly constant \teff (as seems approximately true for the masses considered here; \citet{Baraffe02}).  Then, with eqn. [1], \tkh $\equiv$ $R$/[d$R$/d$\tau$] can be simply integrated to find \citep{Hartmann98}:

$$ \tau \approx \frac{t_{KH}}{3} \eqno{2} $$

where \tkh is the present Kelvin-Helmholtz contraction timescale\footnote{Which intial radius contraction begins at is immaterial in the following sense: the integration produces a [1/{$R^3$($\tau$)} + 1/{$R^3$($\tau=0$)}] term, in which the 1/$R^3$($\tau$=0) part can be ignored as long as initial radius $R$($\tau$=0) is simply much larger than the current radius $R$($\tau$).  In other words, the contraction timescale decreases very rapidly with increasing radius, so the time spent at the largest radii adds negligibly to the final age.}.  Eqns. [1] and [2] capture the essential physics entering the current evolutionary models during the early, pre-D-fusion contraction phase of low-mass objects, though factors such as the precise treatment of convection can modify the results somewhat (mainly by affecting \teff and its variations during collapse; see \citet{Baraffe02}).  They can thus be used to form a rough, internally consistent estimate of the age errors in the models arising from the choice of initial conditions.

According to the Lyon models, the \teff at which 0.1 \msun, 0.03 \msun and 0.02 \msun objects collapse are $\sim$ 3000K, 2700K and 2500K respectively.  In that case, from eqns. [1] and [2], all three reach log \gv = 3.5 at $\tau$ $\approx$ 0.7 Myr, i.e., about a million years after beginning collapse from some very much larger initial radius.  Moreover, a 0.03 \msun object reaches its D-burning phase (log \gv $\sim$3.75) at $\tau$ $\approx$ 1.5 Myr.  Thus, the 1 Myr isochorone in Fig. 9 should actually run roughly parallel to the \teff axis, at log \gv $\approx$ 3.5, instead of rising to log \gv $\sim$3.75 by 0.03 \msun.  By adopting their stated initial conditions, the Lyon isochrones make very low-mass objects, that are truly about 1 Myr old, appear too young by 1--1.5 Myr.  By $\tau$ $\approx$ 2 Myr, however, the isochrone should no longer be a horizontal line.  Now a 0.1 \msun object has already begun fusing deuterium (indeed, is nearing the end of its D-burning phase) and so remains at nearly the same conditions (log \gv, \teff) as at 1 Myr.  However, 0.03 and 0.02 \msun bodies have contracted to log \gv $\approx$ 3.75 (which is the beginning of D-burning for 0.03 \msun).  Thus, the 2 Myr isochrone should closely resemble the curved Lyon 1 Myr track shown in Fig. 9 (and the 3 Myr isochrone should look like the Lyon 2 Myr track and so on).  In summary, the initial conditions adopted in the Lyon models produce {\it (1)} a general under-estimation of age by $\sim$ 1 Myr, and {\it (2)} a significant over-estimation of gravity (by $\sim$ 0.25 dex) for 1 Myr old objects with mass $\lesssim$ 0.03 \msun.  These results quantify the initial condition effects shown in \citet{Baraffe02}, and are vital for very low-mass, very young (1 -- few Myr old) objects.  

However, the above age (or gravity) offsets can account only marginally for our GG Tau results, and not at all for our Upper Sco ones.  The gravities we derive for Ba and Bb are log \gv $\approx$ 3.375 and 3.125 respectively, while in Paper II, we derive masses of $\sim$ 0.12 \msun and 0.03 \msun respectively.  Our formulae above then imply an age of $\sim$ 1 Myr for Ba, but only 0.2 Myr for Bb.  Even accounting for our uncertainty of $\pm$0.25 dex in gravity (and the corresponding factor of 1.8 in mass) can push the age of Bb up to only 0.5 Myr.  Furthermore, the $\sim$ 1 Myr age from Ba is probably only a lower limit, since it has begun fusing deuterium by 1 Myr and thus remains at the same gravity and \teff, without contracting, till $\sim$ 2 Myr\footnote{Fig. 9 indicates that D-burning has not yet begun in Ba, given its log \gv and \teff.  However, our mass for Ba (0.12\msun; Paper II) is higher than Fig. 9 implies; this is simply because we (mostly) find a given mass to be cooler (by $\sim$200K) than the Lyon models suggest (Paper II).  If Ba is at 0.12\msun, it should indeed be in the D-fusion phase by an age $\sim$1 Myr (as seen by simply sliding Ba to the left by 200K in Fig. 9).}.  Thus, even under the best of circumstances, our gravity determinations yield an age difference of at least 0.5 Myr between Ba and Bb.  It is not clear whether this age difference is admissible in current star / brown dwarf formation scenarios, for objects in the same system.  

The situation is even more extreme in Upper Sco.  For the objects with derived gravity of 3.75--4.0 (see Fig. 9), the ages inferred from eqns. [1] and [2] are $\sim$ 3--5 Myr, in good agreement (within 1 Myr) with both the isochronal ages implied by the Lyon tracks (Fig. 9) and the expected age of Upper Sco.  However, we derive log \gv $\approx$ 3.25 for USco 128 and 130, and a mass of only $\sim$ 9--14 \mj; this yields an age of merely $\sim$0.3 Myr for these two objects.  Even including the $\pm$0.25 dex and factor of 2 uncertainty in our gravity and mass respectively (which gives a mass upper limit of $\sim$ 20--30 \mj, similar to the value found by AMB00 for USco 128), still implies an age of at most $\sim$ 0.7 Myr (i.e., close to that of GG Tau Bb).  Thus the age differential with the rest of Upper Sco seems to be of order 3--5 Myr.  

These results can be summarized thus.  A simple \tkh prescription, which encapsulates the evolutionary model calculations at early times (in the absence of D-fusion), implies that all objects with mass $\lesssim$ 0.1 \msun should have log \gv $\gtrsim$ 3.5 dex by an age $\gtrsim$ 1 Myr.  However, our lowest gravity bodies, presumably of age $\gtrsim$ 1 Myr, have log \gv lower than or marginally consistent with 3.5 dex.  This produces age discrepancies in our sample that increase with estimated cluster age.  If our lowest gravities are accurate, then somehow the contraction of these objects must be slowed down in an as yet unaccounted for manner.  

One possibility is through accretion.  As shown by \citet{Hartmann97}, strong accretion can retard the contraction, if the thermal energy of the accreted material can provide the luminosity otherwise supplied through gravitational contraction.  However, this requires that the incoming matter not radiate away all its excess energy, before it has a chance to become well-mixed into the interior and contribute to the object's internal energy.  Even assuming that such mixing can occur efficiently in our objects, one still requires that accretion is either ongoing or has ended only in the very recent past.  Otherwise, since the Kelvin-Helmholtz timescales at the lowest gravities we find are very short (\tkh $\sim$ 0.5 Myr), the objects would already have contracted to higher gravities.  However, GG Tau Bb, USco 128 and USco 130 do not show any overt signs of accretion.  For Bb, \cite{White02} can only find an upper limit to any accretion rate, at 10$^{-10.5}$ \msun{yr$^{-1}$}.  Similarly, the high-resolution spectra of USco 128 and 130 do not indicate any substantial ongoing accretion (JMB02); USco 128 also does not have any detectable millimeter emission, ruling out more than a few Earth masses of surrounding dust, and thus (presumably) any massive accretion disk \citep{Klein03}.  On the other hand, at least two components of the GG Tau system, Aa and Ab, have substantial accretion disks, and thus a small amount of ongoing accretion cannot be ruled out for Bb.  Similarly, USco 128 does have a significant near infrared excess indicating some disk material \citep{jayaetal03}; while 130 does not show any NIR excess, the difficulty in detecting such excess in brown dwarfs suggests that an accretion disk is not completely implausible in this case either.  Consequently, while there seems to be only a slim chance that contraction in these objects has been halted through accretion, the possibility cannot yet be eliminated; it needs to be checked through further observations.


A second, perhaps more attractive possibility is that deuterium-burning begins earlier than the Lyon models anticipate.  Imagine that for masses $\lesssim$ 0.1 \msun, D-fusion really starts somewhere in log \gv $\approx$ 3.25--3.5 range: consistent with the lowest log \gv values we derive, and lower by 0.25--0.5 dex than the D-burning gravities implied by the Lyon models for these masses.  Then the positions of our Taurus and Upper Sco objects on the \teff-gravity plane (Fig. 9) can easily be understood, without doing violence to the previously estimated ages of these clusters or requiring any age-spread within each cluster.  We first outline how this scenario would explain our results, and then discuss its likelihood.  

Consider GG Tau first.  We suggest that the system is indeed $\sim$ 1--2 Myr old, and D-fusion has begun in both Ba and Bb.  Then, since the fusion timescale at $\sim$ 0.03 \msun is close to 5 Myr, the age derived for Bb from a \tkh formulation is only a lower limit on its true age.  Its position on the \teff-gravity plane actually indicates the conditions at which fusion is initiated; Bb will continue to remain there until deuterium is exhausted.  The fusion timescale for Ba ($\sim$ 0.1 \msun), however, is only about 1 Myr, so it may either be coming to the end of its D-burning phase or just past it.  Its position then indicates either its fusion conditions, or a slight subsequent contraction over $<$1 Myr since the end of fusion\footnote{E.g., if D-fusion in Ba begins at log \gv$\sim$3.25, it would take Ba $\sim$ 0.5 Myr to gravitationally contract to that point; it would remain there, burning D, for another 1 Myr, and then contract to log \gv $\sim$ 3.5 (consistent with our derived log \gv) after a further 0.5 Myr, bringing its total age to $\sim$ 2 Myr (as expected for Taurus).}.  

Similarly, we suggest that our higher-mass Upper Sco objects have all finished burning deuterium at $\sim$ log \gv of 3.25--3.5 and rapidly contracted subsequently to their observed gravities by the age of Upper Sco, USco 104 has just finished its fusion and contracted slightly less in the remaining time, and USco 128 and 130 are still in their D-burning phase.  With regard to timescales, this scenario is possible because: {\it (1)} the age of Upper Sco is expected to be $\sim$ 5 Myr; {\it (2)} the D-burning timescale very quickly increases with decreasing mass, going from $\lesssim$ 5 Myr for masses $\gtrsim$ 0.03 \msun to more than 50 Myr for objects at the D-burning limit ($\sim$ 0.012 \msun) - this allows higher-mass brown dwarfs to deplete deuterium over the lifetime of Upper Sco while lower-mass ones continue fusion; and {\it (3)} the contraction timescale at log \gv of 3.5--3.25 is very short, so even if D-fusion happens at these gravities, objects that have finished fusion will very rapidly reach higher gravities (starting at log \gv=3.25--3.5, it only takes $<$2 Myr to contract to log \gv = 3.75, and $<4$ Myr to reach log \gv = 4.0).  Notice that a decrease in gravity, in going from objects more massive than about 0.03 \msun to less massive ones, is actually seen in the 10 Myr isochrone in Fig. 9, for exactly the same reasons we describe here: the higher masses have finished D-burning and subsequently contracted somewhat, while the lower masses are still stuck in the D-fusion phase.  However, the gravity change is much less dramatic than in our proposed scenario, simply because the contraction timescales are much longer at the higher gravities (compared to those in our scenario) at which D-fusion occurs in the Lyon models.  

Of course, this hypothesis can be contemplated only if USco 128 and 130 are above the D-burning minimum mass.  We derive $\sim$ 9--14 \mj for them (Paper II), which is very close to the current theoretical D-fusion boundary of 12 \mj (as \citet{Saumon96} point out, the value of this limiting mass has proved rather robust in the face of various theoretical improvements over the years).  Moreover, our mass errors are about a factor of 2, so it is very plausible that USco 128 and 130 are capable of initiating Deuterium fusion.  
  
The more important issue is whether current theoretical uncertainties can accomodate D-fusion at a lower gravity.  Given the highly complicated nature of ion/electron screening in the D-fusion calculations, especially at the lowest masses \citep{Saumon96}, as well as remaining uncertainties in the interior structure (also exacerbated with decreasing mass), this scenario does not seem infeasible, and is worth examining in future evolutionary models.

We also point out a crucial prediction of our hypothesis that D-fusion might occur at lower-than-expected gravities.  Objects with mass below the D-fusion boundary should not halt in their contraction due to nuclear burning; {\it consequently, their gravities should be much higher, at ages upto a few tens of Myrs, than those of the lowest mass brown dwarfs still in their fusion phase}.  As a result, there should be a sharp trough in the \teff-gravity plane near the D-burning limit in young clusters, with objects at \teff lower than the trough-value {\it certain} to have planetary masses.  The situation is somewhat analogous to the lithium-depletion boundary in the brown dwarf regime.  This would be an exciting development if true.  While the spectral signatures of deuterium are very hard to detect directly with current technology \citep{Chabrier00b}, the spectral signatures of low/high gravity are much easier to decipher (as we have demonstrated in this paper); this would offer a clean way of identifying planetary mass objects in young clusters.  Of course, this possibility needs to be carefully checked through further observations and D-fusion calculations, as noted above.  

\citet{Baraffe02} also examine uncertainties in the treatment of convection.  Basically, at very low gravities (log \gv $\lesssim$ 3.5), the deep atmosphere has extended super-adiabatic layers (due to the suppressed production of \hydmol, and hence lower \hydmol CIA opacity, at such gravities).  As a result, the atmospheric $P$-$T$ profile becomes very sensitive to the precise value of the mixing-length parameter $\alpha$.  This in turn affects the evolutionary models, which use the structure of the deep atmosphere as an outer boundary condition.  At higher gravities, \hydmol is produced efficiently, convection becomes essentially adiabatic, and the dependence on $\alpha$ becomes negligible.  As noted earlier, our synthetic spectra (for which the deep atmosphere is an {\it inner} boundary condition) use $\alpha$=2, which agrees with the value indicated by the latest 3-D hydrodynamical simulations, while the Lyon models shown use $\alpha$=1.  However, \citet{Baraffe02} have illustrated the effect of using $\alpha$=2 atmospheres.  At first glance, the attendant changes in the evolutionary tracks do not seem to improve the discrepancy in gravity between our measured values and the predicted ones for the coolest objects, since the influence of the mixing-length parameter is seen to mostly vanish by an age $\gtrsim$ 1 Myr.  However, \cite{Montalban03} have pointed out shortcomings in the mixing-length treatment of convection adopted in the Lyon89/00 models, arising from super-adiabaticity effects.  Montalban et al. have so far examined only hotter, more massive objects (\teff $>$ 4000K), so the situation in the cool, low-mass regime is unclear.  Moreover, as we point out above, the results of \cite{Baraffe02} suggest that super-adiabatic layers do not last much later than $\sim$1 Myr in our targets.  Nevertheless, Baraffe et al.'s calculations were undertaken within the mixing-length framework.  It remains to be seen whether the effects pointed out by Montalban et al., when applied to the $\lesssim$ 1 Myr evolutionary stage of our targets (when super-adiabaticity {\it is} important, even in the Baraffe et al. models), can produce changes to the evolutionary tracks that continue to be evident at ages greater than a million years.  Certainly, at the gravities we find for our coolest targets, super-adiabatic layers are expected.  These issues must be considered in detail in future evolutionary models.  

Finally, we point out that the Lyon98/00 tracks make use of an older generation of synthetic spectra than we employ here, with quite significant differences in opacity.  In general, this should have a relatively small effect on the isochrones, since the time-evolution of \teff and luminosity is expected, in these low masses, to be a very weak function of opacity ($L$($t$) $\propto$ ${{\kappa}_R}^{{\sim}1/3}$, \teff($t$) $\propto$ ${{\kappa}_R}^{{\sim}1/10}$ ; Burrows and Liebert 1993).  Nevertheless, it would be fruitful to construct new evolutionary models based on the latest opacities, to quantify any resultant change in the isochrones.  

\section{Conclusions}
The primary conclusion of this paper is that it is quite feasible to derive fairly precise (0.25 dex) gravities in low mass stellar and substellar objects. This can be accomplished with spectra of moderate S/N and high resolution (which generally require 8-m class telescopes), along with the most advanced model atmosphere and spectral calculations. If one performs this for objects whose radii can be found (this generally requires reasonably precise distances and photometry), then masses can be found well enough to distinguish planemos, brown dwarfs, and stars (see Paper II). Along with the gravity, we obtain a good effective temperature ($\pm$50K), which facilitates the conversion of a luminosity to a radius.

The basic reason for our success is that TiO is a spectral diagnostic that is strongly dependent on temperature, and weakly dependent on gravity (and with the opposite gravity dependence of our atomic line diagnostics). Our gravity diagnostics are both subordinate and resonance neutral alkali lines (of potassium and sodium). They have both a temperature and gravity dependence, and yield degenerate solutions in the two parameters. We therefore rely on TiO to break the temperature degeneracy, and the alkali lines then provide a gravity. We note that there are a number of other alkali lines that we have not yet employed (other lines of potassium and sodium, plus lines of rubidium, cesium, and lithium), which cover a range of wavelengths and have different strengths for different stellar temperatures. In the L dwarfs, a substitute for the TiO temperature diagnostic must be found (since TiO has condensed into dust). The metal hydride molecules (FeH and CrH) are the obvious candidates. Properly developed, our methodology should work well over the range of substellar objects from mid-M through L (which encompass a range of masses from the planetary to stellar domains, depending on age). Indeed, once theoretical tracks have been properly computed and calibrated, one might be able to directly infer age as well (in principle).

Having applied our methodology to very young, low-mass objects in the Upper Scorpius and Taurus star-forming regions, we find {\it (1)} a good agreement between our gravities and the theoretical ones for most of our sample, but {\it (2)} a significant discrepancy between the two for our coolest targets.  We show that even without detailed comparison against model spectra, the data themselves suggest a sharp fall-off in gravity in the coolest objects.  We have examined various processes and synthetic spectral uncertainties which might lead us to infer erroneous gravities (e.g., dust, cool spots, metallicity variations, and inadequacies in the model treatment of collisional broadening), and found that they are unlikely to give rise to the spectral effects we see: real gravity variations are indicated. Theoretical tracks would interpret this range of gravities as a range of ages, and we provide several arguments against that interpretation. The alternative is that the theoretical tracks are allowing the very low-mass objects to contract too quickly; the gravities we measure for them are lower than predicted. We discuss various ways in which this could happen, and suggest remaining uncertainties in the model treatment of accretion, deuterium-burning and/or convection are most likely responsible.  This means that the interpretation of masses and ages from isochrone analysis in young clusters is problematic at the moment for very low-mass bodies. Nonetheless, we show in Paper II that the interpretation of the faintest of these objects as low-mass brown dwarfs or planetary mass objects looks correct.

Finally, we point out that our spectral analysis is carried out in the optical, but it could potentially be accomplished at other wavelengths as well.  In particular, the near-IR would be very interesting to explore.  Such investigations have already begun: \cite{Doppmann03a} and \cite{Doppmann03b} present a similar high-resolution study in the near-IR for higher mass, hotter PMS objects, while \cite{Gorlova03} examine near-IR PMS substellar spectra, albeit at low-resolution.  A high-resolution analysis of cool, very low-mass PMS objects  would be extremely fruitful, especially for probing the youngest, most extincted bodies; we have recently embarked on a project to accomplish this.  

\acknowledgments
We would like to acknowledge the great cultural and religious significance of Mauna Kea for native Hawaiians, and express our gratitude for permission to observe from atop this mountain.  We would also like to express our thanks to the Keck Observatory staff, who have made possible, and successful, our observations over the last several years.  We would like to thank Russel White and Lynne Hillenbrand for kindly supplying two of the spectra used in this paper.  We would also like to thank Russel White, Lee Hartmann, Gilles Chabrier and Isabelle Baraffe for illuminating discussions on PMS evolution, and a constant readiness to help.  This work was supported in part by NSF grants AST-0098468 to G.B., and AST-0205130 to R.J.  S.M. would like to acknowledge the support of the SIM-YSO grant, for funding his postdoctoral research.   

\clearpage
\appendix
\section{Dust Effects}
Here we investigate dust formation in our PMS sample, by comparing their spectral properties to those of field M dwarfs of similar type.  The M dwarfs chosen are Gl 406 (M6V), VB 8 (M7V), VB 10 (M8V) and LHS 2924 (M9V), all well-known spectral standards for their types.  In these stars, later type is supposed to correlate well with decreasing \teff (e.g., \cite{Jones97}).  In order to isolate \teff-related dust effects as much as possible, we have chosen stars likely to have very similar metallicities and gravities.  Spectral analyses indicate that all four are consistent with solar metallicity (\cite{Leggett00}; \cite{Leggett98}; \cite{Schweitzer96}).  Gravity is constrained by the fact that these are all Main Sequence dwarfs belonging kinematically to the old disk population (\cite{Leggett92}; corresponding, at least in a statistical sense, to ages $\gtrsim$ 3 Gyr).  This largely precludes their being young low-gravity objects still in the contraction phase.  The BCAH98 and CBAH00 evolutionary models (expected to be quite accurate for older MS dwarfs, notwithstanding any uncertainties in the PMS stage; e.g., \cite{Segransan03}) indicate that masses between $\sim$ 0.08 and 0.3 \msun (which adequately cover the expected range for M6-M9 MS dwarfs) all have log \gv $\sim$ 5.0, spanning at most $\sim$ 0.25 dex, by a few Gyrs.  The rough gravities derived so far from spectral syntheses of these objects (\cite{Leggett00}; \cite{Leggett98}; \cite{Schweitzer96}) are also consistent with minimal gravity variation.  At any rate, since \teff is expected to be primarily responsible for determining the spectral type ordering of these dwarfs, and since their spectral effects considered below do change quasi-monotonically (increasing, then decreasing) with type, we can safely assume we are tracking \teff-related effects, and ignore any small gravity or metallicity variations in our qualitative analysis here.  Finally, we can ignore \vsini effects: VB8, VB10 and LHS 2924 all have comparable \vsini (6--10 \kms), and we have artificially broadened Gl 406 (intrinsically $<$ 3 \kms) to 7 \kms to match the others.
   
In Fig.10, we show three spectral regions in the M dwarfs - the triple-headed TiO band at $\sim$ 8440\AA~ (same region used for our PMS \teff analysis), the TiO bands at $\sim$ 7050\AA, and the \na doublet at $\sim$ 8200\AA~ (same region used for PMS gravity analysis).  Three effects are immediately appearent: {\it (1)} the 8440 TiO bands increase monotonically in strength with decreasing temperature, from M6V to M8V, and then reverse strength at still lower \teff, becoming weaker by M9V; {\it (2)} the 7050 TiO bands show a similar behaviour, but reverse strength at an earlier spectral type (M8V), i.e., at a higher \teff, than the 8440 bands; and {\it (3)} the \na doublet behaves analogously to the 7050 bands, reversing strength by M8V and becoming weaker with still later type (lower \teff).  These effects can be understood through \teff-dependent dust formation.  Initially, no dust forms, while the abundance of TiO and \na increases with decreasing \teff, making their lines stronger with later type.  As \teff continues to decrease, dust begins to form, simultaneously depleting TiO from the gas phase (since TiO becomes sequestered in grains) and increasing the continuum opacity.  Consequently, the TiO absorption bands and the \na doublet now weaken with later type.  Moreover, dust opacity increases with shorter wavelength.  As a result, its effects are manifested first (ie, at an earlier type, or higher \teff) in the bluer TiO bands at 7050\AA, and only later in the redder bands around 8440\AA.  Qualitatively, one might expect \na, occurring at an intermediate wavelength (8200\AA), to be affected at an intermediate spectral type / \teff; our analysis here shows that it follows the behaviour of the 7050\AA~ TiO bands more closely.  Finally, since dust appears in the 7050 bands before it does in the 8440 ones, the magnitude of the reversal in the former is larger than in the latter at a given spectral type / \teff: e.g., by the time the 8440 bands reverse around M9V to become weaker than at M8V, the 7050 bands are already {\it much} weaker than at M8V (Fig. 10).    

Turning now to our PMS objects, we have already illustrated the behaviour of their 8440 TiO bands and \na doublets in earlier sections.  In Fig. 10, we plot the trend in their 7050 TiO bands as \teff decreases.  Note that our \teff are derived from the 8440 bands assuming no dust, so decreasing \teff actually means increasing strength in the latter bands.  We see that, as our derived \teff declines, the 7050 band-strength either remains roughly constant, or increases (the reason for this variable increase in strength is discussed shortly).  Most importantly, this TiO band shows {\it no} reversal with lower \teff, even in those objects with anomalously weak \na.  This implies that our ordering of objects by \teff is correct, and that dust is unimportant in our PMS sample and cannot explain our \na results.

To see this, consider USco 55, 104 and 130.  As discussed in the main text, the \na doublet in USco 130 is slightly weaker, and in 104 much weaker, than in USco 55; meanwhile the 8440 TiO bands increase steadily in strength from USco 55 to 130 (Figs. 6 and 8).  This is surpring at first, since deeper 8440 bands usually imply lower \teff, and \na should strengthen with decreasing \teff.  However, our M dwarf analysis above shows that burgeoning dust opacity can make the \na lines weaker with declining \teff, even as the 8440 bands continue to deepen; the question is whether this is what is happening in our PMS objects.  We do {\it not} think so, based on the 7050 TiO results.  The M dwarf spectra reveal that the latter bands should show a strong reversal when dust produces a reversal in \na, even if no dust effect is apparent in the redder 8440 bands.  The fact that we see no weakening at all of the 7050 bands in USco 104 and 130 compared to 55 - these bands are about the same strength as in USco 104 and 55, and stronger in 130 - argues convincingly that dust is not responsible for the \na behaviour in our PMS objects.  

Finally, it is also interesting that the 7050 bands in USco 104 are about the same strength as in 55.  Of course, this may simply be because their \teff are so similar: perhaps the 7050 bands are not as sensitive to a small \teff change as the 8440 ones.  However, we suggest that gravity may play a role as well.  To illustrate this clearly, we compare the behaviour of 104 to that of USco 109 (same \teff as 104, but significantly higher gravity; Table 1), and USco 112 (significantly higher \teff than 104 and 109, but same gravity as 109; Table 1).  In Fig. 10, we see that the 7050 bands in 109 are deeper than in 112, as expected fom their \teff difference (derived from the 8440 bands).  However, while the 7050 bands in 104 are also slightly deeper than in 112, the difference is smaller than between 109 and 112 (Fig. 10).  Since 109 and 104 appear to have exactly the same \teff (i.e., indistinguishable 8440 bands; not shown), this suggests that there is some other factor distinguishing the two; their disparity in gravity is the obvious choice (they also differ by $\sim$8\kms in \vsini, but tests show that this has negligible effect on the depth of the strong 7050 bands).  In fact, our synthetic spectra also indicate that lower gravity, at a given \teff, causes the 7050 bands to weaken slightly.  In other words, while the behaviour of the observed 7050 TiO bands strongly suggests no dust formation in our PMS sample, it does support the gravity variations we have deduced earlier. 
 
\section{GG Tau B:  \teff and Gravity}
GG Tau is a PMS quadruple system in the Taurus-Aurigae star-forming region (D = 140pc).  It is a heirarchical system composed of two binaries: GG Tau A and GG Tau B.  The two components of the first, Aa and Ab, are separated by $\sim$ 0''.25 ($\sim$ 35 AU); the two components of the second, Ba and Bb, are separated by 1''.48 ($\sim$ 207 AU).  The separation between the two binary systems is $\sim$ 10''.1 ($\sim$ 1414 AU).  A circumbinary disk is also observed around GG Tau A (e.g., \cite{Guilloteau99}).

WGRS99 find spectral types of K7$\pm$1 and M0.5$\pm$0.5 for Aa and Ab respectively, and M5$\pm$0.5 and M7$\pm$0.5 for Ba and Bb.  High-resolution spectra of the GG Tau A system show low levels of continuum excess, or veiling, indicating some ongoing accretion in this binary (\cite{Gullbring98}).  In GG Tau B, however, WGRS99 detect no accretion-induced continuum excess in either component; they infer an upper limit of 10$^{-9}$ \msun{yr$^{-1}$} for the accretion rate in both Ba and Bb.  They also find a \vsini of 9$\pm$1 \kms in Ba, and 8$\pm$1 \kms in Bb.  In a later detailed analysis, \cite{Luhman99} find M5.5 for Ba and M7.5 for Bb.  Still more recently, White \& Basri (2002) derive spectral types of M6 and M7.5, and \vsini of 8.1$\pm$0.9 and 6.6$\pm$2.0 \kms, for Ba and Bb respectively.  They also put more stringent upper limits on any possible accretion, at $<$ 10$^{-9.8}$ and $<$ 10$^{-10.3}$ \msun{yr$^{-1}$} for Ba and Bb. 

The 4 components of GG Tau are very likely to be coeval, as WGRS99 argue.  Using this fact, they estimate the age and component masses of the system, in the following manner.  Extinctions are found for the 4 components using the derived spectral types, and comparing the optical and NIR colors to those of field dwarf spectral standards.  Luminosities are then determined by applying bolometric corrections to the dereddened photometric magnitudes.  \teff, however, are {\it not} precisely derived a priori, but assumed to be intermediate between those of giants and dwarfs of the same spectral types (the rationale for this is discussed shortly; note also that WGRS99 adopt the dwarf \teff scale of \cite{Leggett96}, which is noteworthy for our discussion further on).  At the K7--M0.5 types of Aa and Ab, the dwarf and giant \teff are nearly identical, so the \teff of these two components is essentially `fixed'.  At the later types of Ba and Bb, however, the dwarf and giant scales diverge, and thus the \teff of these two is only constrained to a rather broad range of $\sim$ 400K.  WGRS99 then compare these luminosity and \teff estimates to the predictions of the theoretical evolutionary tracks constructed by various groups.  In all the different models examined, a unique isochrone can be found that simultaneously fits Aa and Ab; i.e., all the models successfully predict the coevality of these two stars.  Not all, however, are capable of also fitting Ba and Bb with the same track.  That is, using the track that fits Aa and Ab, many of the models predict \teff for Ba and/or Bb that are oustside the assumed `acceptable' range (delimited by the dwarf and giant \teff scales).  WGRS99 use this fact to distinguish between the different models.  They find that the BCAH98 models (Lyon98/00 in our nomenclature) are to be preferred, since these predict \teff for Ba and Bb that are commensurate with the \teff range assumed for the latter.  Thus, the BCAH98 models are consistent with the coevality of all 4 components of GG Tau, but {\it only within the broad \teff range assumed for Ba and Bb}.  Indeed, WGRS99 finally adopt, for Ba and Bb, the \teff implied by the BCAH98 models themselves; this is essentially assuming coevality to determine \teff (\cite{Luhman99} explicitly uses this procedure to define a PMS \teff scale), under the sole constraint (apart from known luminosity) that the \teff so derived falls within a wide range of acceptable values.  Whether BCAH98 models actually predict coevality when \teff is better constrained is not addressed by WGRS99 (as distinct from our analysis, which attempts to rigidly restrict \teff before comparing to model tracks).    

The WGRS99 comparison to BCAH98 models yields an age of 1.5 Myr for the GG Tau system, and masses of 0.78$\pm$0.10, 0.68$\pm$0.03, 0.12$\pm$0.02 and 0.044$\pm$0.006 \msun for Aa, Ab, Ba and Bb.  For Ba and Bb, the \teff implied by the BCAH98 tracks are 3050K and 2820K, and the gravities are log \gv = 3.53 and 3.65, respectively\footnote{The gravities implied by BCAH98 are not explicitly quoted in WGRS99.  However, they can easily be calculated from the luminosities, \teff and masses supplied.  We have also checked the values we quote here against the Lyon98/00 models directly, for the masses found by WGRS99.}.  WGRS99 note that the Ba is then $\sim$ 40K hotter than an M5 dwarf, while Bb is $\sim$ 140K hotter than an M7 dwarf (using dwarf temperatures from Leggett et al. (1996; hereafter, L96)). 

We have derived \teff and log \gv for Ba and Bb using HIRES optical spectra kindly supplied to us by R. White.  Unfortunately, the spectrograph setting used by White exludes the \na doublet, as well as the blue component of the \pot doublet, though it does include the TiO order we have used for our Upper Sco sample.  However, in the very high S/N spectrum obtained by R. White, even the single component of the \pot doublet in the spectra turns out to be capable of adequately constraining the gravity, while the TiO order, as before, is an excellent \teff indicator.  In Figs. 11, we plot our synthetic fits to the data.  The fits obtained are evidently remarkably good.  For GG Tau Ba, we are unable to distinguish between log \gv=3.25, \teff=2750K and log \gv=3.50, \teff=2800K.  We thus adopt the average, log \gv=3.375, \teff=2775K, with errors of $\pm$ 0.25 dex and $\pm$ 50K respectively.  Similarly for Bb, we obtain log \gv = 3.125 $\pm$ 0.25 dex, \teff = 2575 $\pm$ 50K.

The comparison between our results and the Lyon98/00 models have been presented in \S 5.3.  Here we briefly discuss our differences with the \teff found by WGRS99.  Our temperatures for both Ba and Bb are $\sim$ 200K lower than the WGRS99 results.  This makes our values somewhat lower than even the dwarf \teff that WGRS99 quote (from L96), for their spectral types for Ba and Bb.  This is noteworthy, since WGRS99 assume that PMS temperatures should be intermediate between the giant and dwarf scales (i.e., somewhat higher than dwarf values).  Let us first state what the rationale for this assumption is.  In general, it is found that the optical spectra of mid- to late M PMS stars are well-reproduced by the averaged spectra of dwarfs and giants of the same spectral type (for earlier M-types, dwarf spectra alone seem better; see \cite{White02}).  In fact, types are often {\it assigned} to these cool PMS objects on the basis of the giant and dwarf spectral types that in combination reproduce the PMS spectrum (this is the procedure adopted by WGRS99 for GG Tau Ba and Bb; see e.g., WGRS99, \cite{Luhman99}, \cite{White02}).  As a rough estimate, therefore, it is perhaps not unreasonable to suggest that the temperatures of such PMS objects, just like their spectra, are intermediate between those of giants and dwarfs\footnote{The other rationale is that an intermediate PMS scale agrees with the BCAH98 predictions (\cite{Luhman99}).  However, while this makes such a scale very useful for comparing different samples to the same set of tracks, it is clearly not germane to our analysis, since we wish to probe the intrinsic veracity of the tracks.}.  In this light, the lower-than-dwarf \teff we derive may appear troubling.

However, recent revisions in the field dwarf \teff scale, as well as in the spectral types of Ba and Bb, remove any cause for anxiety.  As mentioned earlier, Ba and Bb have now been pushed to somewhat later types (M7.5 for Bb, and M5.5--6 for Ba), compared to the WGRS99 estimates.  Concurrently, Leggett et al. (2000) (hereafter, L00) have derived \teff for a large sample of disk and halo M dwarfs (M1-M6.5), by fitting low-resolution optical to near-infrared spectra with (low-resolution versions of) the same models we have used.   For M5 dwarfs, they find \teff $\sim$ 2900K, for M5.5, $\sim$ 2800K, and for M6-6.5, $\sim$ 2600K.  Their sample does not include any dwarfs M7 or later\footnote{It does include one metal-poor (M/H $\sim$ -0.5) M7 sub-dwarf; at 2900 K, it is as hot as many M5 objects.  Low metallicity usually leads to a higher \teff; this sub-dwarf is thus not representative of M7-7.5 dwarfs.}.  In general however, M7-7.5 dwarfs should be cooler than the M6-6.5 ones.  The L00 result for the M6-6.5 types then implies \teff $\lesssim$ 2500K for M7-7.5; this is also supported by the results of \cite{Leggett01}.  These values are likely to be more accurate than those WGRS99 compile from L96; the new models used by L00 fit the observed SEDs much better than the substantially older models in L96.  Importantly, these new temperatures for mid- to late M dwarfs are 100--200K lower than the L96 estimates.  If these newer values are compared to our \teff for Ba and Bb, then Bb is $\sim$ 100K hotter than an M7-7.5 dwarf.  In fact, this is very similar to what WGRS99 find by comparing the BCAH98 predictions to the older L96 dwarf scale.  This is not surprising: while our \teff for Bb is $\sim$ 200K cooler than the WGRS99 one, the dwarf temperature scale has also been lowered by roughly the same amount.  Similarly, since L00 find M5.5-M6 dwarfs to be at $\sim$ 2800--2600K, our \teff for Ba (2800K) is at least comparable to the new dwarf scale (which is what WGRS99 found using the older L96 scale), and perhaps $\sim$ 200K hotter.  Thus, our \teff estimates for both Ba and Bb are completely consistent with a PMS temperature scale intermediate between that of dwarfs and giants, once recent improvements in the dwarf \teff scale, and in the spectral type of Ba and Bb, are taken into account.

\newpage
\plotone{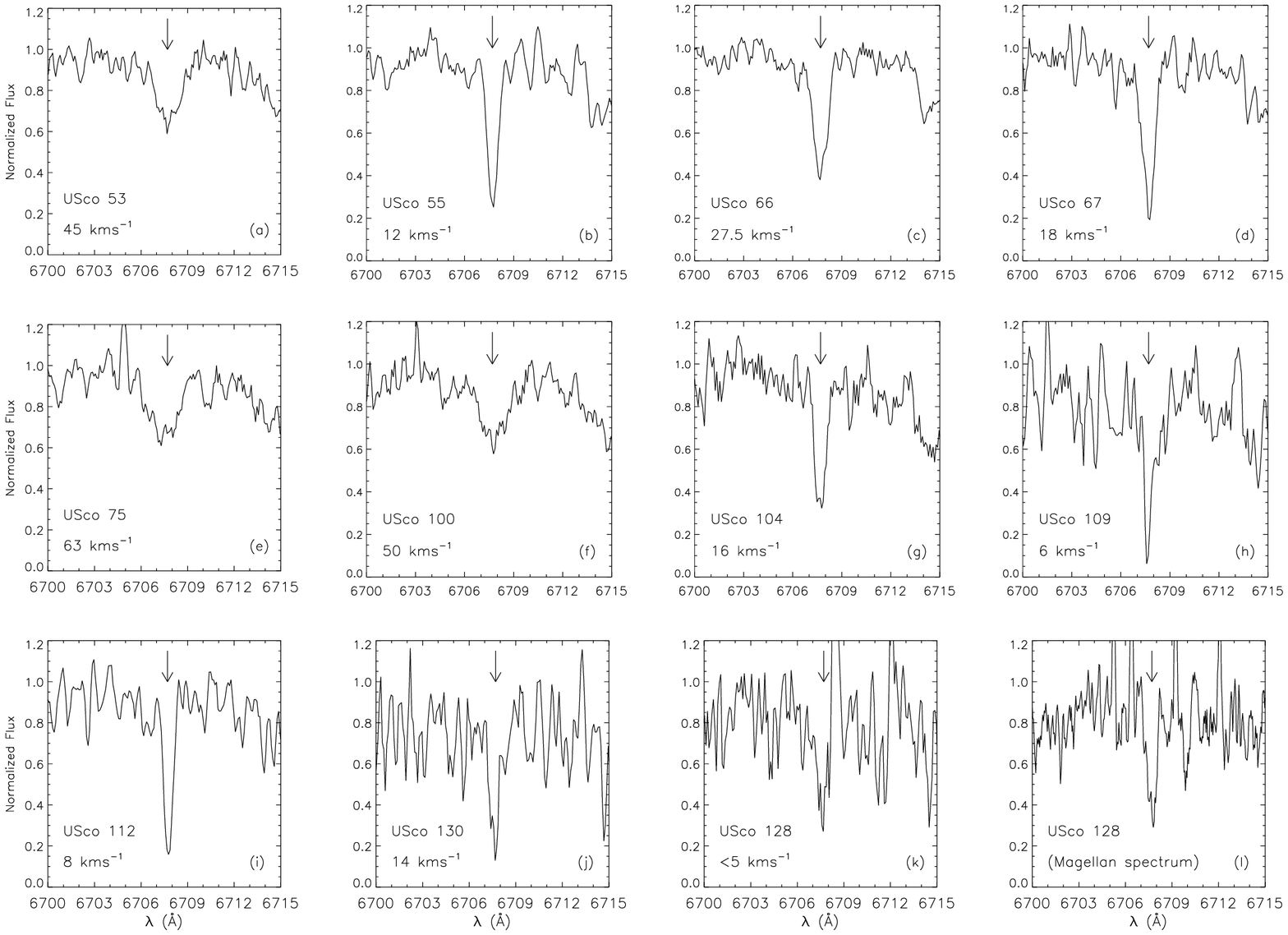}
\figcaption{\label{fig1} Lithium in our HIRES Upper Sco sample.  This and following plots: all spectra smoothed by 3-pixel boxcar, unless otherwise noted.  Object names and \vsini are noted.  LiI is confidently detected in all, except USco 128 (panel {\it k}), where the detection is at lower S/N.  However, it is clearly present in the high-resolution Magellan spectrum ({panel \it l}).}

\plotone{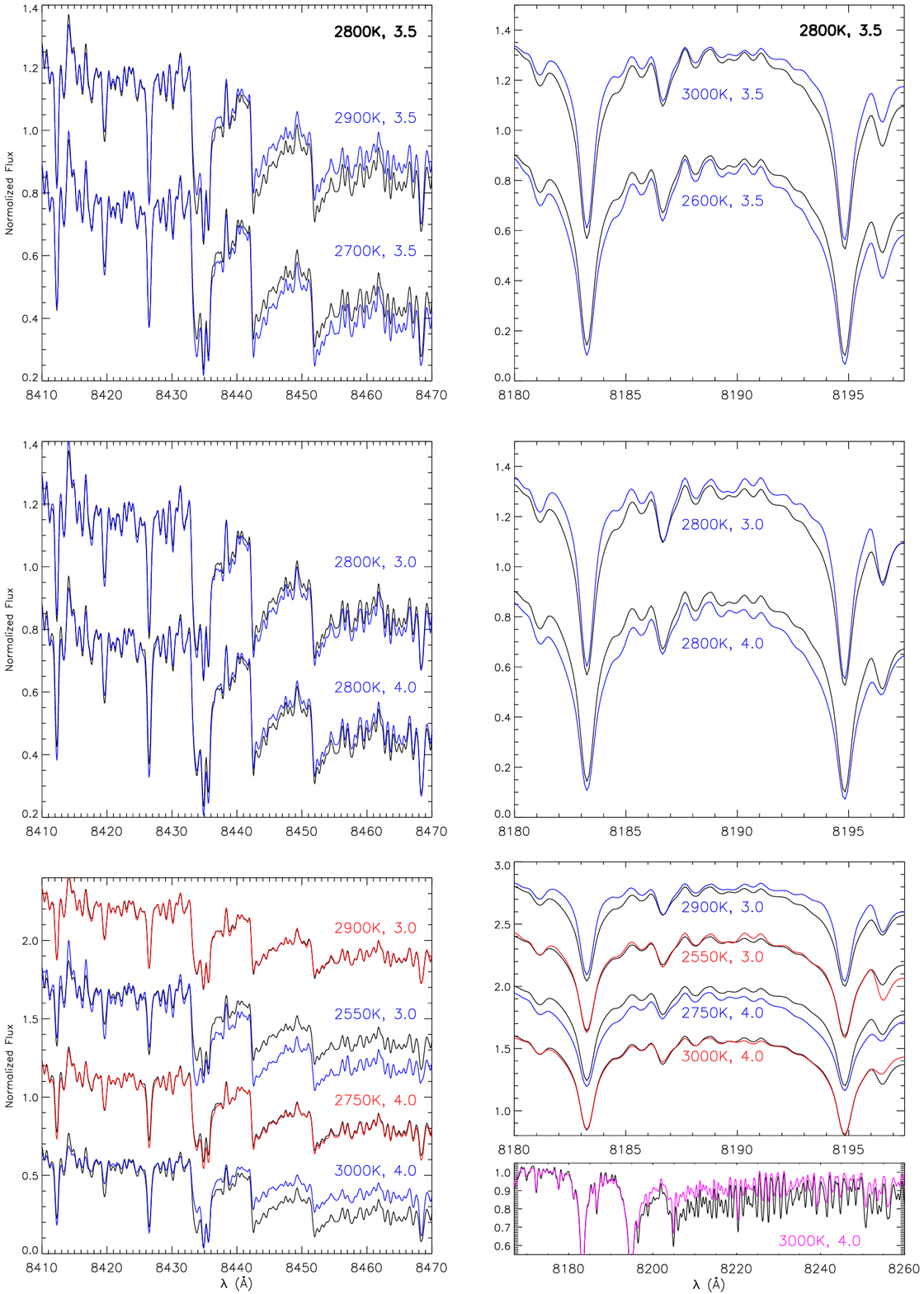}
\figcaption{\label{fig2} Dependence of TiO and \na on \teff and gravity, in the model spectra.  In all panels of both left and right columns, the {\it black} spectrum is the synthetic spectrum at \teff=2800K, log \gv=3.5.  We use this as a comparison template, to illustrate changes in TiO and \na with varying \teff and gravity.  This and following plots: All spectra normalized by a multiplicative factor; multiple spectra in a plot are offset by additive constants for clarity.  {\it Top left:}  The strong dependence of the TiO bandheads (esp. at 8442\AA~ and 8452\AA~) on \teff; decreasing \teff makes the bands deeper.  {\it Top right:}  The strong dependence of the \na doublet on \teff; decreasing \teff makes the doublet broader.  {\it Center left:}  The weak dependence of TiO on gravity; at a given \teff, the TiO band-strengths remain almost constant over a 1 dex range in log \gv.  {\it Center right:}  The strong dependence of the \na doublet on gravity; at a constant \teff, increasing gravity makes the doublet broader.  {\it Bottom right, large panel:}  Degeneracy between \teff and gravity effects on the \na doublet; a \teff increase of 200K almost exactly compensates for a gravity increase of 0.5 dex, enabling a given \na profile to be fit by various \teff / gravity combinations.   {Bottom left:} TiO comparisons for the same gravities and temperatures shown for \na in the large bottom right panel.  While fits can be obtained to \na with various \teff/gravity combinations, the same combinations do not give simultaneous good fits to TiO; demanding simultaneous fits to TiO and \na thus allows us to quickly converge on the correct \teff and gravity (see text).  {\it Bottom right, small panel:}  The \teff dependence of the continuum around the \na doublet; while the doublet itself shows a \teff-gravity degeneracy, the continuum is much more sensitive to \teff alone.  Varying \teff produces continuum mismatches. Thus, while the 3000K, log \gv=4.0 and 2800K, log \gv=3.5 spectra are very similar in the profile of the \na doublet (bottom right, large panel), the two spectra are clearly different in the surrounding continuum (bottom right, small panel).  Demanding a simultaneous good fit to the \na doublet and to the surrounding continuum provides an additional constraint on \teff and gravity. }

\plotone{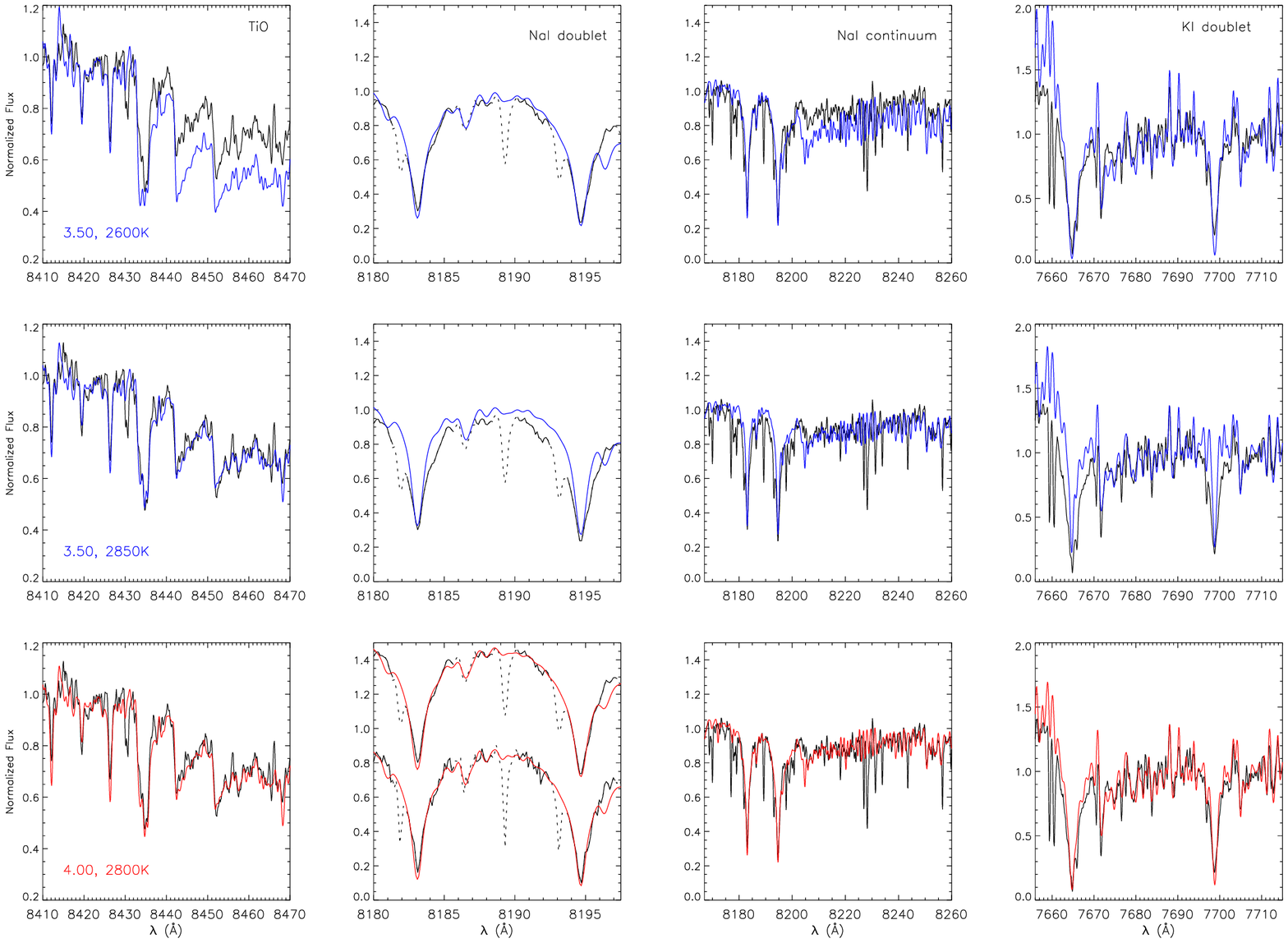}
\figcaption{\label{fig3} Model fits to USco 55.  See text.  Data in black, synthetic spectra in color.  Unacceptable fits in blue, adopted fit in red.  The fit parameters for each row are noted in the first column.  First column shows triple-headed TiO bandhead, second shows \na doublet, third shows entire \na order and fourth shows \pot doublet.  Telluric absorption around the \na doublet shown as dotted lines, in second column.  The sharp absorption lines in the \na continuum, as well as blueward of the \pot doublet, are also telluric.  {\it Top row:} log \gv=3.50, \teff=2600K model gives good fit to \na doublet, but bad fits to TiO and \na continuum (and marginally good fits to the \pot doublet: the predicted red lobe is deeper, and the predicted continuum blueward of the blue lobe much stronger, than observed).  {\it Center row:}  log \gv=3.50, \teff=2850K gives good fits to TiO and the \na continuum, but bad fits to the \na and \pot doublets.  {\it Bottom row:}  log \gv=4.0, \teff=2800K model gives good fits to all four spectral regions shown.  For the \na doublet, we show two fits; both are at the same \teff and gravity, but the upper fit is with smoothed spectra, and the lower one with unsmoothed.  Both fits are equally good, illustrating that noise in the data is not affecting our inferred \teff and log \gv.}

\plotone{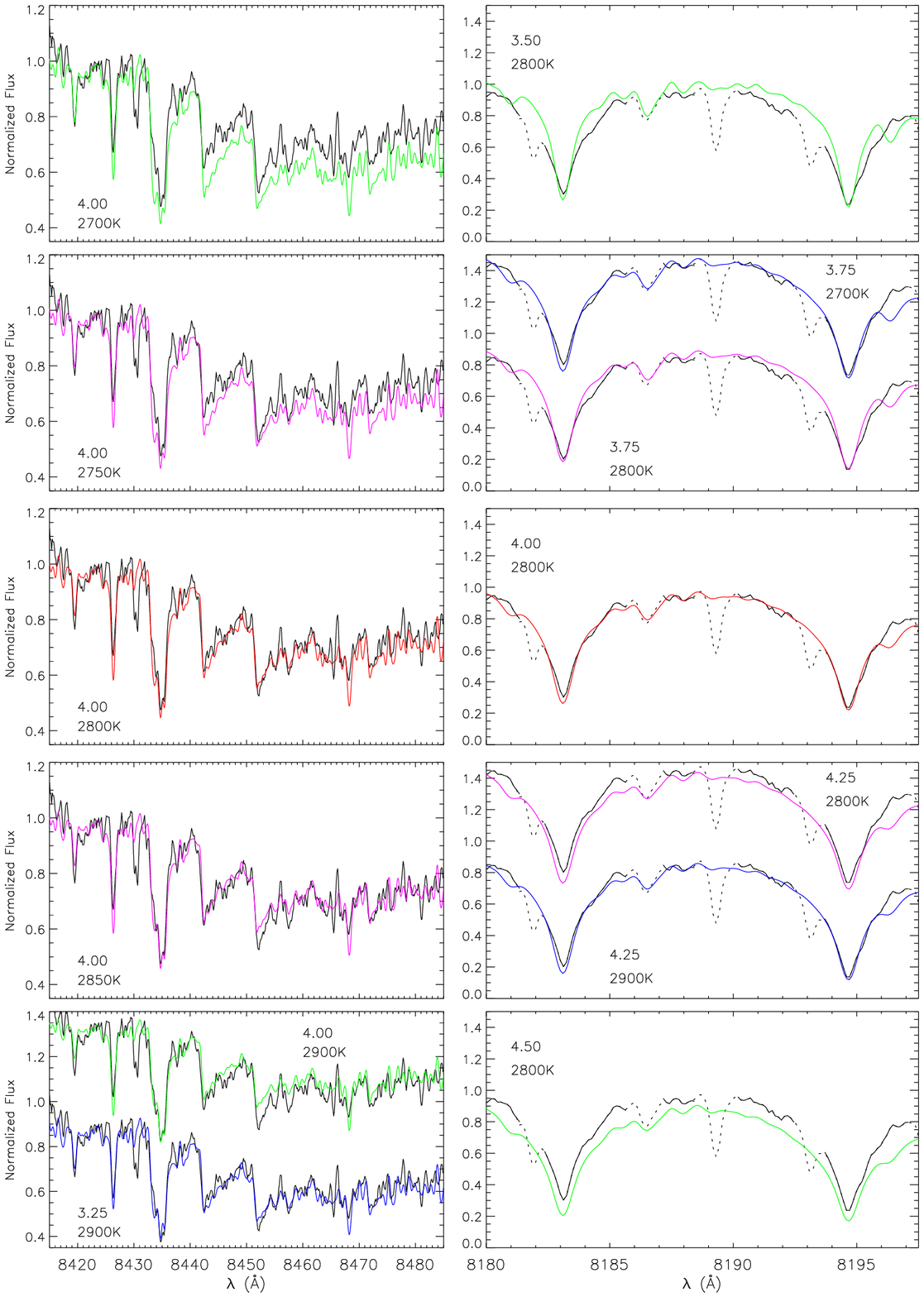}
\figcaption{\label{fig4}  Error analysis in USco 55.  {\it Left column:}  TiO data (black) compared to models (in color) at various \teff and log \gv=4.0.  In red, we show the best fit to TiO (log\gv=4.0, \teff=2800K) obtained in Fig. 3.  In purple, we show models at 2800$\pm$50K.  These fits are slightly worse, either in the bandhead strength or in the continuum redward of 8452\AA~.  In green, we show models at 2800$\pm$100K.  The fits are clearly very poor.  Collectively, this shows that TiO is sensitive to 100K \teff changes, and that $\pm$50K is a good estimate of our \teff error.  {\it Right column:}  \na doublet data (black) compared to models (in color) at various \teff and gravities.  Again, we show in red the best fit to \na obtained earlier (4.0, 2800K).  In purple, we show models at the same \teff, but with log \gv=4.0$\pm$0.25 dex.  These fits are slightly worse.  Better fits are obtained at the same gravities but with \teff=2800$\pm$100K (blue).  However, these \teff are clearly not good fits to TiO (see left column).  Finally, in green, we again show models at 2800K, but with log \gv=4.0$\pm$0.50 dex.  These are obviously poor fits.  All this shows that, once \teff is fixed by the TiO fits, \na is very sensistive to 0.50 dex gravity changes; $\pm$0.25 dex is a good estimate of our gravity error. }

\plotone{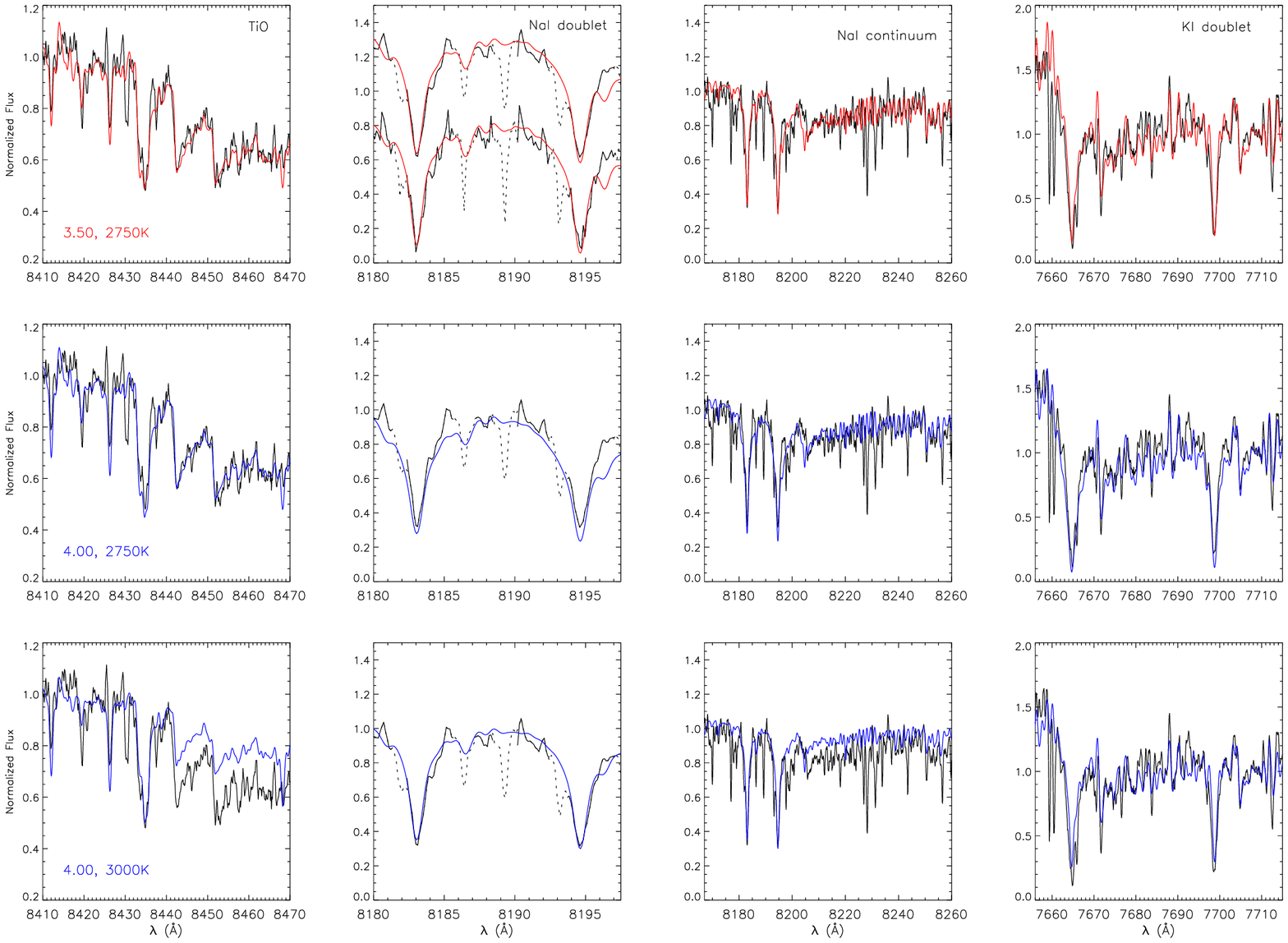}
\figcaption{\label{fig5} Model fits to USco 104; otherwise same as Fig. 3.  See text.  {\it Top row:}  log \gv=3.50, \teff=2750K model gives the best fit to the data, in all the spectral regions shown.  For the \na doublet, we show two fits; both are at the same \teff and gravity, but the upper fit is with smoothed spectra and the lower one with unsmoothed.  Both fits are equally good, showing that noise in the data is not affecting our derived parameters. {\it Center row:}  log \gv=4.0, \teff=2750K is a good fit to TiO and the \na continuum, but is significantly stronger than the observed \na doublet, and slightly stronger than the \pot doublet.  {\it Bottom row:}  log \gv=4.0, \teff=3000K is a good fit to the \na and \pot doublets, but a bad fit to the TiO bandheads and \na continuum. }

\plotone{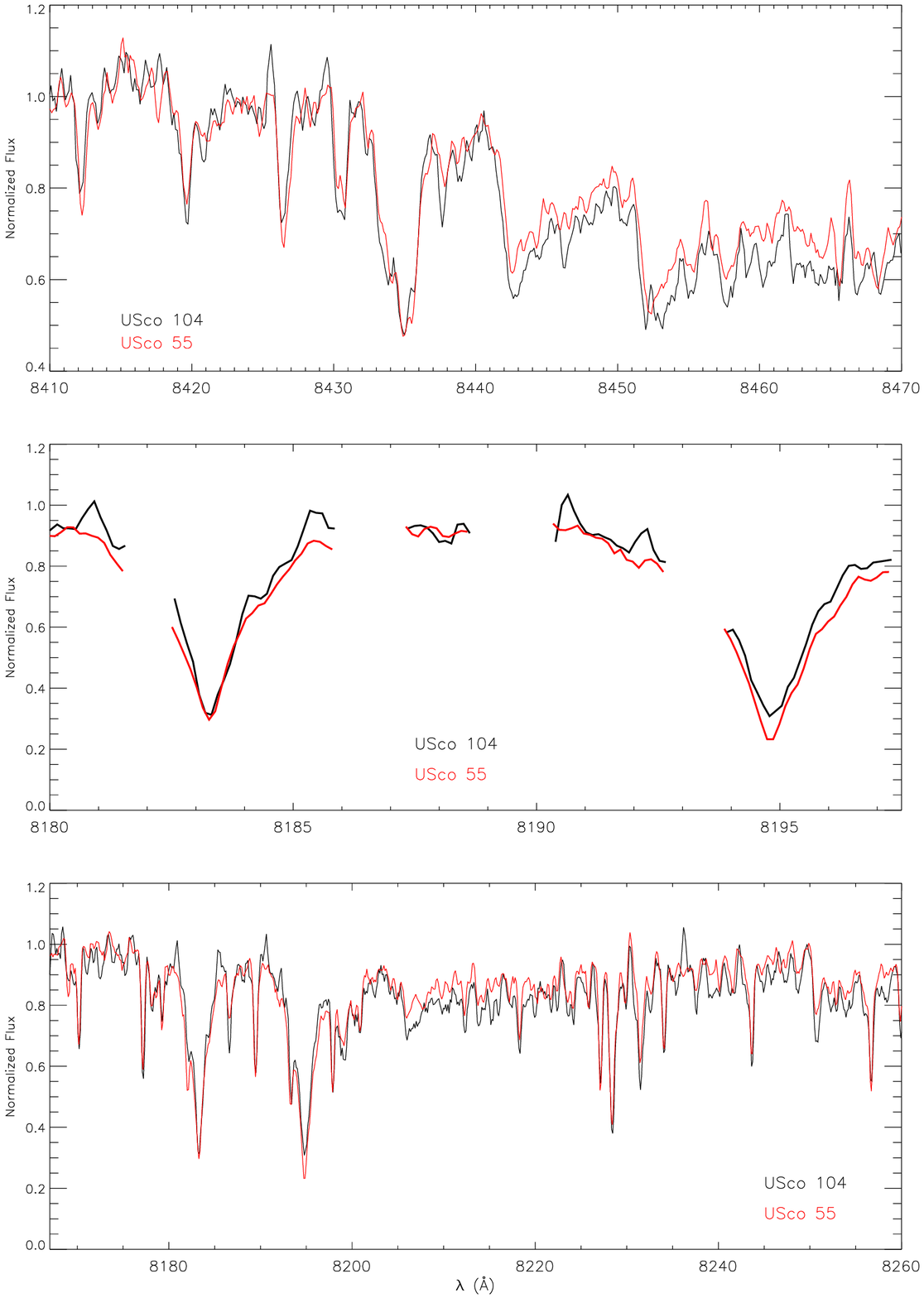}
\figcaption{\label{fig6} USco 104 compared to USco 55.  See text.  USco 104 in black, USco 55 in red.  {\it Top:}  TiO bandheads in USo 104 compared to 55; the two are almost identical, implying very similar \teff.  Close examination shows that the USco 55 bandheads at 8442 and 8452\AA are slightly shallower than in 104; this is in agreement with our finding USco 55 to have slightly (50K) higher temperature.  {\it Center:}  \na doublet in USco 104 compared to 55.  Telluric lines are excluded for clarity.  Both spectra normalized to a continuum region outside the plot (but see bottom plot).  \na in USco 104 is appreciably weaker, though its \teff is lower, thus implying a lower gravity than in 55.  {\it Bottom:}  Entire \na order in USco 104 compared to 55.  The continuum in the two objects overlap very well, implying no normalization errors.  The general shape and slope of the continuua also agree very well, implying very similar \teff. Notice that the TiO bandheads in USco 55, at $\sim$ 8205 and 8250 \AA~, are slightly weaker, indicating that USco 55 is in fact very slightly hotter than USco 104 (as indicated by the TiO bandheads in the top panel as well).}

\plotone{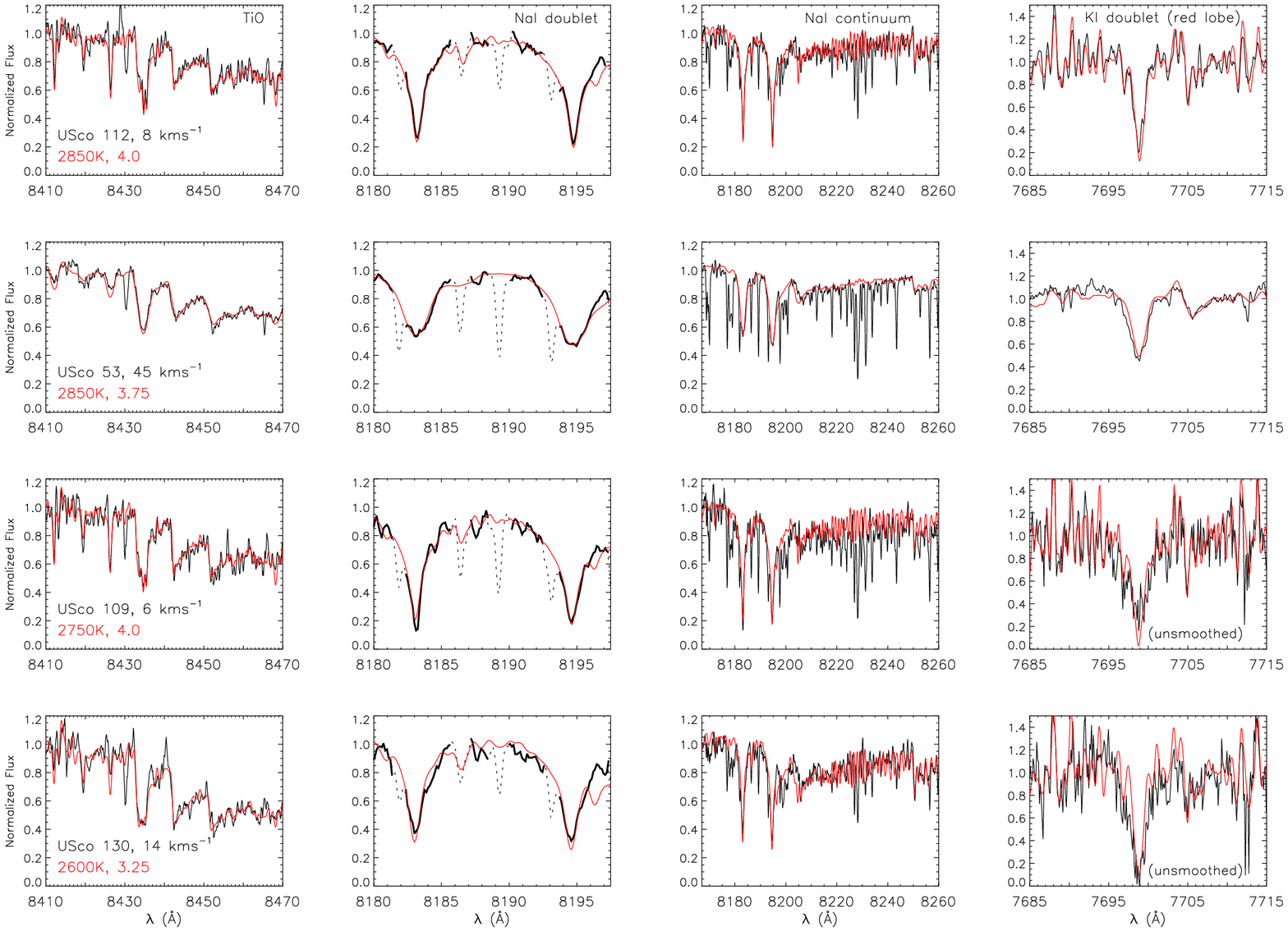}
\figcaption{\label{fig7} Best fits to some of our Upper Sco objects.  First column shows TiO bandheads, second shows \na doublet, third shows entire \na order, fourth shows only red lobe of \pot continuum (blue lobe is generally telluric-contaminated).  Form top to bottom: USco 112, 53, 109 and 130.  The model parameters are noted in the plots.  Note the excellent fits obtained to all the spectra regions shown with the same model parameters.  Note also how the models exclude the sharp telluric lines in the \na order.  The fit obtained to \pot in USco 130 is not very good, due to noisy data in this spectral region; the noise is illustrated by plotting the unsmoothed \pot data (and model) for this object.  See text.  }

\plotone{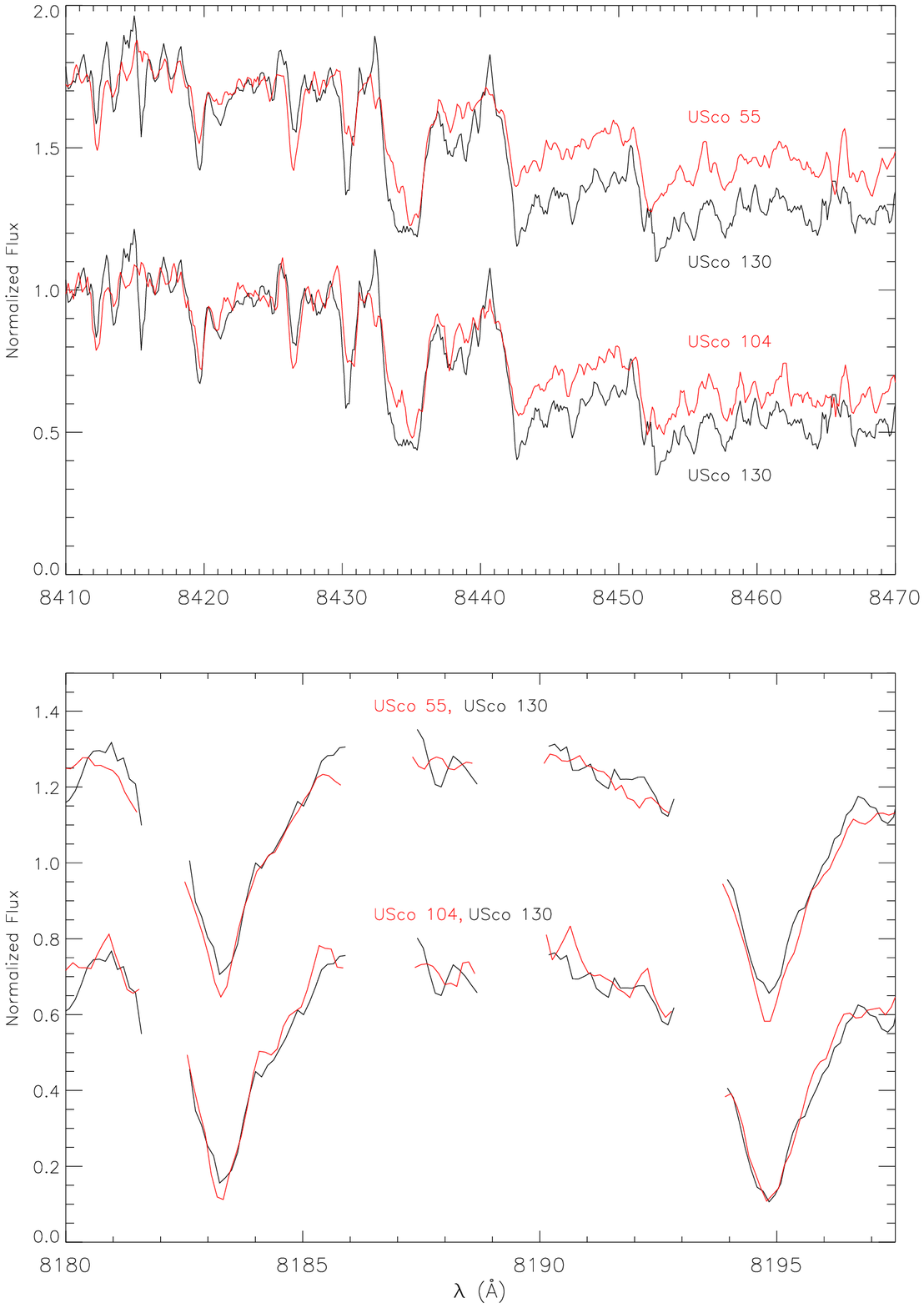}
\figcaption{\label{fig8} Usco 130 (black) compared to USco 55 and 104 (both red).  TiO bandheads in 130 (esp. at 8442 and 8452\AA~) are significantly deeper than in 55 and 104, implying a lower \teff ({\it top panel}).  Despite this, its \na doublet is very similar in strength to that of 55 and 104 ({\it bottom panel}).  This can only happen if USco 130 has a significantly lower gravity than in the other two objects. }

\plotone{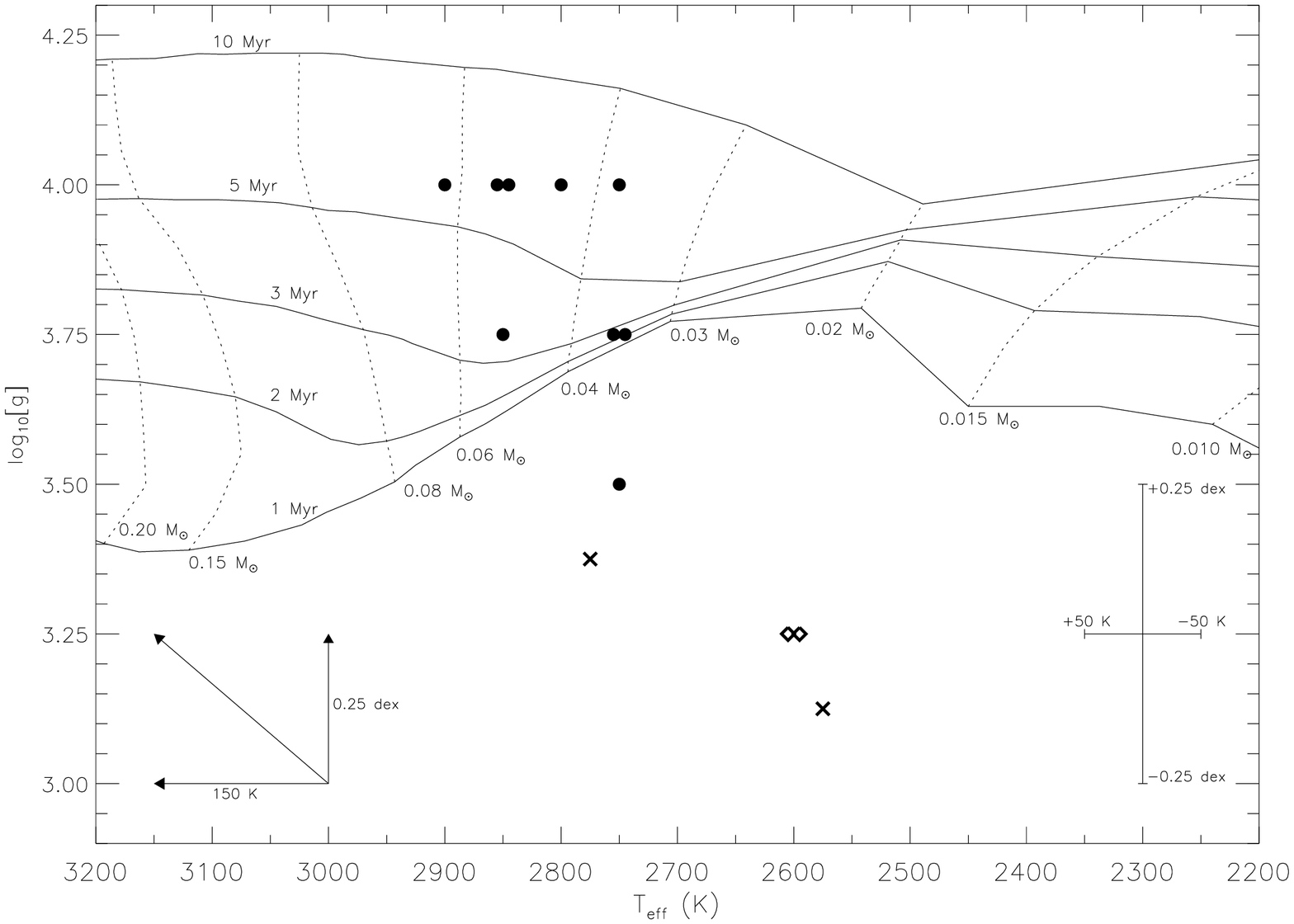}
\figcaption{\label{fig9} \teff versus gravity for our Upper Scorpius and Taurus sample, compared to the Lyon98/00 predictions.  All Upper Sco members are shown as filled circles, except the two coolest ones, USco 130 and 128 (diamonds).  GG Tau Ba and Bb are the crosses.  Objects with the same derived \teff and gravity have been slightly separated in \teff, for clarity.  Solid lines are isochrones in the Lyon98/00 model, at ages from 1 to 10 Myr.  Dotted lines denote the evolutionary tracks for various masses in the Lyon98/00 model, from 0.20 to 0.01 \msun.  The error bars on our measurements of \teff ($\pm$50K) and gravity ($\pm$0.25 dex) are depicted at the bottom right.  USco members with \teff $\gtrsim$ 2750K agree fairly well with the Lyon98/00 5 Myr locus.  Cooler objects, however, appear much younger.  Similarly, GG Tau Ba is not too far from the expected 1 Myr locus, but Bb seems much younger.  The horizontal and vertical arrows, in the bottom left, indicate the change in \teff and log \gv (150K and 0.25 dex respectively) required if any object is afflicted by very large, cool spots (50\% areal coverage, 500K cooler than surrounding photosphere; see \S 4.3.2); the diagonal arrow shows the combined change in position.  }

\plotone{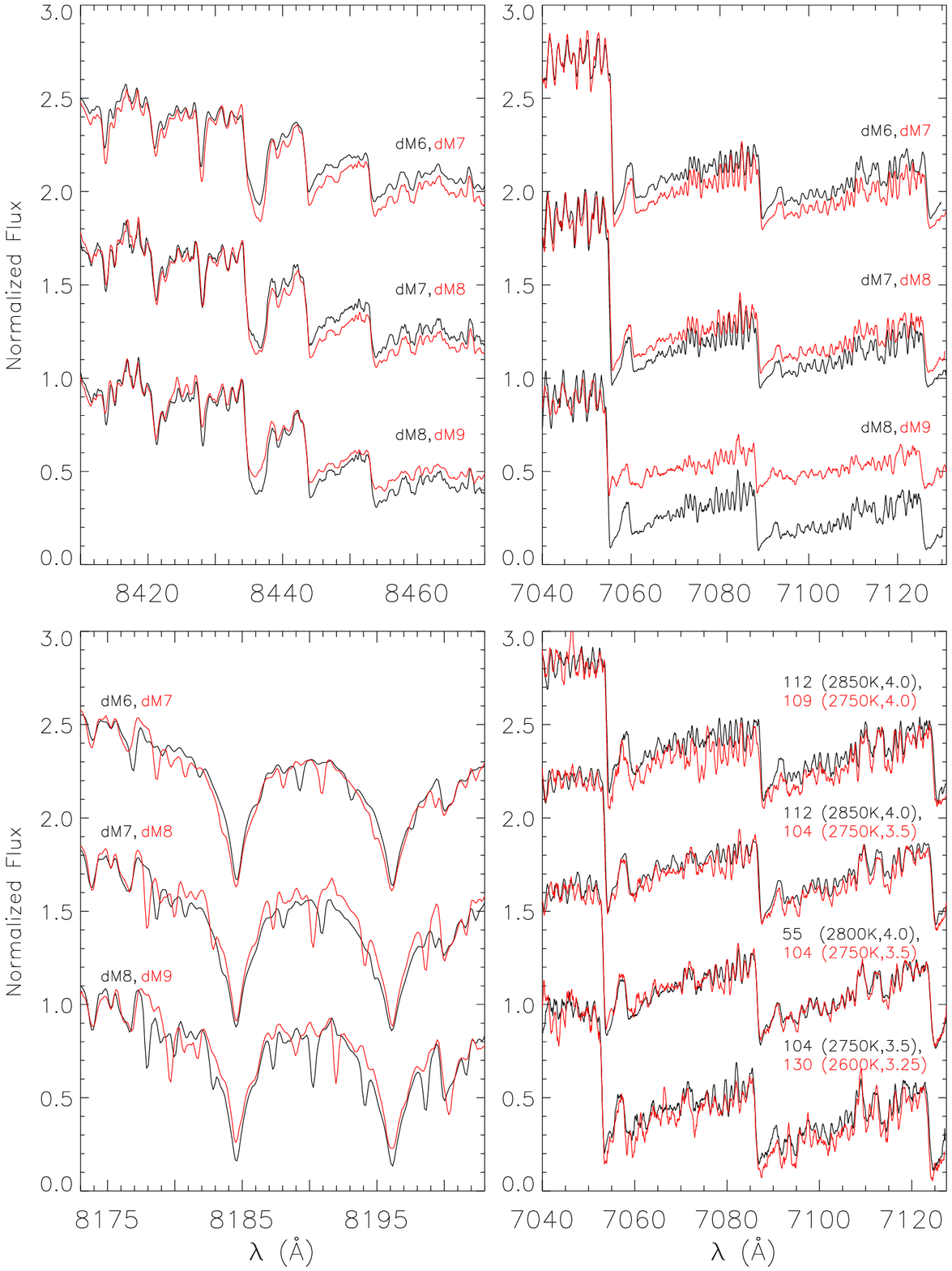}
\figcaption{\label{fig10} Dust effects in field M dwarfs, compared to USco sample.  {\it Top left:} Comparison of 8440\AA~ TiO bands in a spectral sequence of M dwarfs.  {\it Top right:} Comparison of 7550\AA~ TiO bands in the same M dwarfs.  {\it Bottom left:} Comparison of \na doublet in the same M dwarfs.  {\it Bottom right:} Comparison of 7050\AA~ TiO bands in our PMS sample.  With increasing dust opacity, the M dwarfs first show a reversal in the strength of the 7050 TiO bands and \na doublet, and, at later spectral types (i.e., lower \teff), in the 8440\AA~ TiO bands.  No such reversal in the 7050 bands is apparent in our PMS sample.  See \S 4.3.1 and Appendix A. }

\plotone{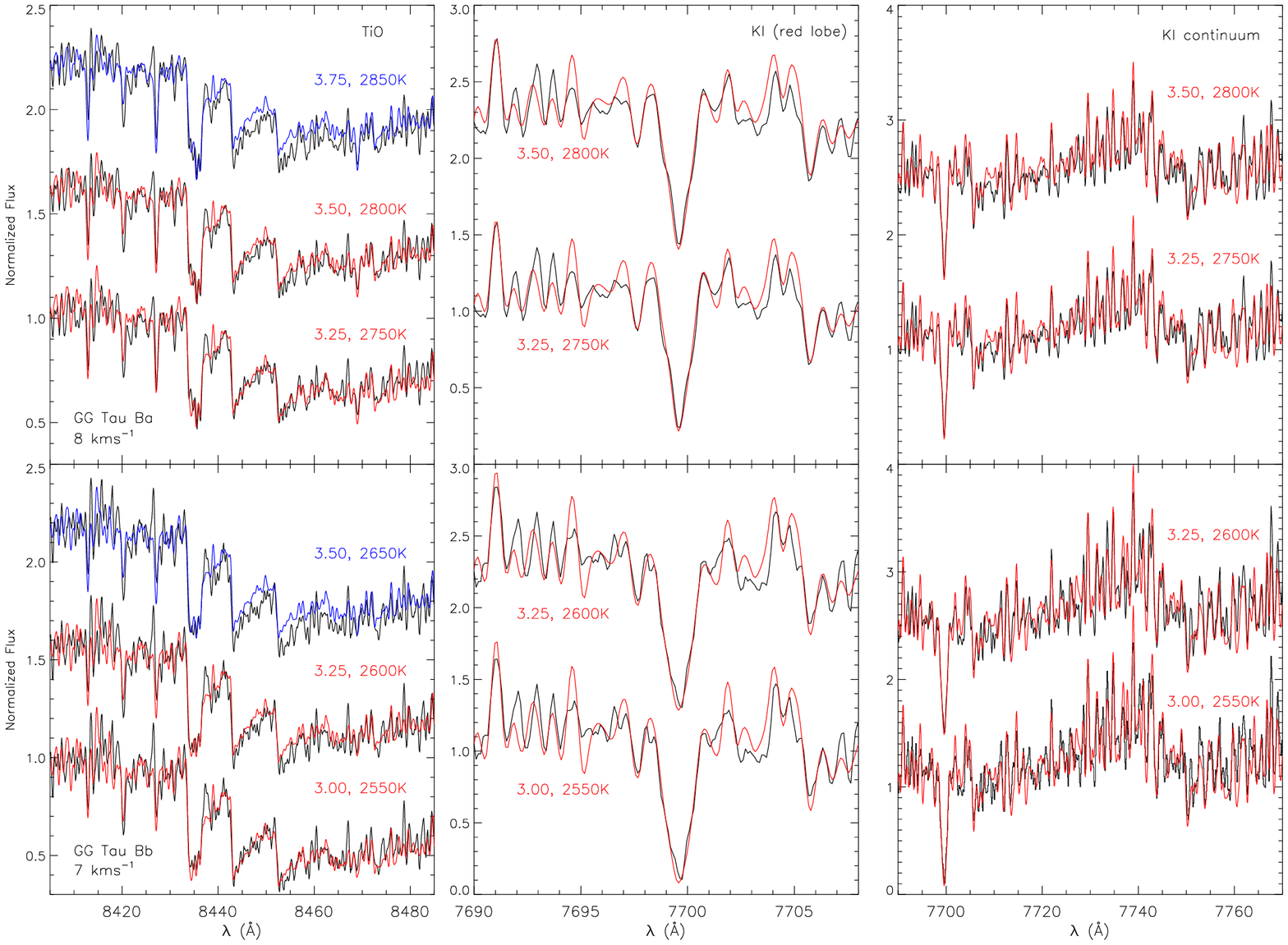}
\figcaption{\label{fig11} Model fits to GG Tau Ba and Bb.  The top row shows fits to Ba, the bottom row to Bb.  Data is in black, models in color.  Unacceptable fits are in blue, adopted fits in red.  The first column shows the TiO bandheads, the second column the red lobe of the \pot doublet, the third column the entire \pot order.  {\it Top row:}  Both log \gv=3.25, \teff=2750 and log\gv=3.50, \teff=2800K models are good fits to all the Ba spectral regions shown.  We adopt the average: log \gv=3.375, \teff=2775K.  We also show that the log \gv=3.75, 2850K model is a poor fit to TiO (though it is a good fit to the \pot doublet; not shown).  Thus 3.75, 2850K can be excluded.  {\it Bottom row:}  Same, for Bb.  Both log \gv=3.00, \teff=2550K and log\gv=3.25, \teff=2600K models are good fits to all the Bb spectral regions shown.  We adopt log \gv=3.125, \teff=2575K.  We also show that log \gv=3.50, 2650K is a poor fit to TiO (though it is a good fit to the \pot doublet; not shown).  Consequently 3.5, 2650K can be excluded.}

\clearpage

\begin{deluxetable}{lccccc}
\tablecaption{\label{tab1} Derived Parameters}
\tablewidth{0pt}
\tablehead{
\colhead{name\tablenotemark{a}} &
\colhead{\vsini\tablenotemark{b}} &
\colhead{\teff\tablenotemark{c}} &
\colhead{log \gv\tablenotemark{c}} &
\colhead{SpT\tablenotemark{d}}  \\
 &(\kms) &(K) & & \\} 
                           
\startdata

USco 66        & 27.5$\pm$2.5   &2900  & 4.00 &M6  \\
USco 112       & 8$\pm$2    &2850  & 4.00 &M5.5  \\
USco 75        & 63$\pm$5   &2850  & 4.00 &M6  \\
USco 53        & 45$\pm$2.5   &2850  & 3.75 &M5  \\
USco 55        & 12$\pm$3   &2800  & 4.00 &M5.5  \\
USco 109       & 6$\pm$2    &2750  & 4.00 &M6  \\
USco 67        & 18$\pm$2   &2750  & 3.75 &M5.5  \\
USco 100       & 50$\pm$3   &2750  & 3.75 &M7  \\
USco 104       & 16$\pm$2   &2750  & 3.50 &M5  \\
USco 128       & $<$5 &2600  & 3.25 &M7  \\
USco 130       & 14$\pm$2   &2600  & 3.25 &M7.5  \\
\\
GG Tau Ba & 8$\pm$2    &2775  & 3.375 &M6  \\
GG Tau Bb & 7$\pm$2    &2575  & 3.125 &M7.5  \\

\enddata
\tablenotetext{a}{Objects within a cluster are ordered by decreasing \teff.  At a given \teff, they are aranged by decreasing gravity.}
\tablenotetext{b}{\vsini values same as in JMB02, except for USco 112, where we find 8 \kms versus 6 \kms in JMB02 (see \S 2.2)}
\tablenotetext{c}{\teff errors are $\pm$ 50K, log \gv errors are $\pm$ 0.25 dex.}
\tablenotetext{d}{Spectral types for Upper Sco objects from AMB00, and for Taurus ones (GG Tau Ba and Bb) from White \& Basri (2002).} 
\end{deluxetable}

\end{document}